%% file: kmsetc.tex
\documentclass[a4paper,12pt,twoside]{article}
\usepackage{amssymb,amsmath,amsthm,amsfonts,latexsym,amscd}
\numberwithin{equation}{section}

\title{KMS, \textsc{etc.}}
\author{Lothar Birke%
\footnote{\texttt{lothar@itp.phys.ethz.ch}, supported
in part by Swiss National Fund}
 and   J\"urg Fr\"ohlich\footnote{\texttt{juerg@itp.phys.ethz.ch}}}
\date{Theoretical Physics\\ETH-H\"onggerberg\\CH-8093 Z\"urich}

\newcommand{\Anaught}{\stackrel{\,\,\,\circ}{\mathcal A}}

\newtheorem{lemma}{Lemma}
\newtheorem{thm}[lemma]{Theorem}
\newtheorem{corl}[lemma]{Corollary}

\begin{document}
\maketitle
{\small \textbf{Abstract:} 
A general form of the {\it ``Wick rotation''}, starting from {\it
  imaginary-time Green functions} of quantum-mechanical systems in
thermal equilibrium at positive temperature, is established. 
Extending work of
H. Araki, the r\^{o}le of the {\it KMS condition} and of an associated
anti-unitary symmetry operation, the {\it ``modular conjugation''}, in
constructing analytic continuations of Green functions from real- to
imaginary times, {\it and back}, is clarified.

The relationship between the KMS condition for the vacuum with respect 
to Lorentz boosts, on one hand, and the spin-statistics connection and
the PCT theorem, on the other hand, in local, relativistic quantum
field theory is recalled.

General results on the reconstruction of local quantum theories in
various non-trivial gravitational backgrounds from ``Euclidian
amplitudes'' are presented. In particular, a general form of the KMS
condition is proposed and applied, e.g., to the {\it Unruh-} and the
{\it Hawking effects.}

\medskip

{\it This paper is dedicated to Huzihiro Araki on the 
occasion of his seventieth
birthday, with admiration,
  affection and best wishes.}

}

\section{Introduction and Summary of Results}
The purpose of this paper is to review the {\em general theory of
quantum-mechani\-cal matter in thermal equilibrium} and to describe
some applications of this theory to quantum field theory, in 
particular to theories in some curved space-times. We shall
emphasize the r\^ole played by {\em imaginary-time 
(``temperature-ordered'') Green functions} (TOGF's) in the 
analysis of quantum-mechanical systems in thermal equilibrium,
because the TOGF's are the objects that are most accessible to
analytical studies of such systems based on functional integration;
see~[\ref{G},\ref{FKT}] and references given there, and~[\ref{CFS}]. Many
features of quantum systems with infinitely many degrees of freedom,
such as phase transitions and long-range order, critical behaviour,
strong correlations, etc.\ are encoded into the TOGF's. Nevertheless,
it is the {\em real-time Green functions} (RTGF's) of systems in
thermal equilibrium which are the {\em physical objects}. In order
to calculate e.g.\ the response of such systems to small changes
in the external control parameters, we need to know their RTGF's.

The connection between the RTGF's and the TOGF's of a quantum
system in thermal equilibrium is analogous to
the one between Wightman distributions and Schwinger functions
of a local relativistic quantum field theory (QFT) at zero
temperature, which has been unravelled in the work of 
Osterwalder and Schrader~[\ref{OS}], see also~[\ref{Gl}], building
on a lot of previous, deep work in axiomatic quantum field
theory, see~[\ref{J},\ref{StW}] and references given there: One passes
from TOGF's to RTGF's, and back, by {\em analytic continuation}
in the time variables {\em(``Wick rotation'')}. However,
in contrast to the situation in local, relativistic QFT at
{\em zero temperature}, one {\em cannot} make use of an 
(energy-) {\em spectrum condition}, in order to accomplish the
analytic continuation at positive temperatures. While at 
zero temperature, the Hamiltonian of any reasonable quantum
system is bounded from below, the spectrum of the 
{\em thermal Hamiltonian}, or {\em``Liouvillian''}, of a 
system with infinitely many degrees of freedom at positive
temperature usually covers the {\em entire real axis}. At zero
temperature, the analytic continuation from real to imaginary
time, and back, is based on the fact that if the Hamiltonian
$H$ is a {\em non-negative operator}, then $\exp(izH)$ is
bounded in operator norm by 1, provided $\mathrm{Im}(z)>0$.
At positive temperature, the analytic continuation of RTGF's
in the time variables to the TOGF's, and back, is based on
the {\em Kubo-Martin-Schwinger (KMS) condition}~[\ref{K},\ref{MS}]
known to characterize thermal equilibrium states of quantum
systems. The very formulation of the KMS condition for RTGF's
involves an analytic continuation of RTGF's in one time-difference
variable. An application of the generalized tube theorem then
implies joint analyticity of RTGF's in {\em all} time variables
in a tubular domain containing, as a subset, cyclically ordered
$n$-tuples of {\em imaginary} times. The TOGF's are the 
{\em restrictions} of the analytically continued RTGF's to the 
subset of {\em cyclically ordered imaginary time arguments}.

The main problem studied in this paper is to start from Green
functions (calculated e.g.\ with the help of functional integrals)
which have all the properties of TOGF's --- including an
invariance under cyclic rearrangements of their arguments, which
is the imaginary-time version of the KMS condition --- and prove
that they can be analytically continued in their 
(imaginary-)time arguments back to real times to yield RTGF's 
with all the right properties. Thus, we present a variant
of the {\em Osterwalder-Schrader reconstruction theorem at
positive temperature}.

The reader is right in assuming that this cannot be a new
result. However, while all the elements of our constructions
have appeared in the literature, a complete synthesis does
not appear to have been presented anywhere. It therefore
seems worthwhile to attempt such a synthesis.

The interest of the senior author in these problems goes back
to the first half of the 70's. It was triggered by the work
of Osterwalder and Schrader~[\ref{OS}] mentioned above, Ruelle's
continuation~[\ref{Ru}] of Ginibre's work on reduced density
matrices~[\ref{G}], the classic work of Haag, Hugenholtz and
Winnink~[\ref{HHW}] on KMS states in quantum statistical 
mechanics, Araki's analytic continuation of RTGF's~[\ref{A1}],
and some work of Hoegh-Krohn on thermal field theory~[\ref{HK}].
First results appeared in~[\ref{F1}]. In preparing a course
on statistical mechanics at Princeton University~[\ref{F2}], he
also became familiar with important work of Araki~[\ref{A2}] on
``relative Hamiltonians''. This led to a translation of the 
results of~[\ref{HHW}] to {\em imaginary time}~[\ref{F2}]. A basic
step towards a general (Osterwalder-Schrader type)
{\em reconstruction theorem at positive temperature} was 
undertaken in~[\ref{F3}], with crucial help by E. Nelson.
Subsequently, there was important parallel work by A. Klein
and L. Landau~[\ref{KL1},\ref{KL2}]. However, in their work, use is
made of mathematical structure, in particular of notions from
the theory of random fields, which is not intrinsic to the
general theory of KMS states. It may thus appear to be of
interest to present details of some {\em general results}
on the connection between RTGF's and TOGF's, even though,
informally, they have been known since the late 1970's.

Ever since the work of Bisognano and Wichman~[\ref{BW}],
it has been known that the KMS condition also plays a 
fundamental r\^ole in relativistic QFT at {\em zero}
temperature. The {\em vacuum} is a KMS state for every
one-parameter subgroup of {\em Lorentz boosts}. This
observation is intimately related to (and based on) 
the connection between spin and statistics~[\ref{J},\ref{StW}]
and Jost's general form of the PCT theorem~[\ref{J2}].
This will be briefly recalled towards the end of the
paper from the point of view of an {\em imaginary-time
formulation} of QFT. In particular, the KMS condition
at imaginary time will be seen to be a consequence of 
{\em locality} and of the {\em connection between spin
and statistics} (and conversely!) and to give rise to 
a direct definition of the {\em anti-unitary PCT symmetry
operation}.

The paper is concluded with 
lengthy comments on the imaginary-time
formulation of QFT on some {\em curved space-times}, in
particular the space-time of a Schwarzschild black hole
and de Sitter space. Recalling some general results on 
``virtual representations of symmetric spaces'' proven
in~[\ref{FOS}], it is shown how to reconstruct unitary
representations of the Killing symmetries of space-time.
The KMS condition then yields obvious variants of the
spin-statistics connection and of the PCT theorem and
provides general interpretations of the {\em Unruh-}
and the {\em Hawking effects.}

\textbf{Acknowledgements:} The senior author is deeply
grateful to his scientific grandfathers, fathers and
uncles for having created the atmosphere and the facts
which made considerations like the ones presented in this
paper appear worthwhile and possible. He thanks his collaborators
in work on which this review is based, and, in particular,
E. Nelson and E. Seiler, for all they have taught him. He
is grateful to H. Epstein for very helpful discussions, and
to H. Araki for support and encouragement.

\section{KMS States According to Haag-Hugenholtz-Winnink, and Araki}
\subsection{Finite systems in thermal equilibrium}

Consider a quantum-mechanical
physical system confined to a compact subset of space.
Its time-evolution is generated by a self-adjoint Hamiltonian,
$H$, on the Hilbert space, $\mathcal H$, of pure physical state vectors.
The energy spectrum of $H$ is discrete. Let $Q_1,\ldots,Q_N$ be
self-adjoint operators on $\mathcal H$ representing conserved
quantities (i.e., ``$[H,Q_i]=0$'', $i=1,\ldots,N$) and commuting
with all ``observables'', which are identified with the 
self-adjoint operators in a subalgebra, $\mathcal A$, of the
algebra of all bounded operators on $\mathcal H$. Let
$\mu_1,\ldots,\mu_N$ denote the chemical potentials conjugate
to the conserved quantities $Q_1,\ldots,Q_N$. As recognized
by Landau and von Neumann, the state, 
$\langle(\cdot)\rangle_{\beta,\underline{\mu}}$, of the 
system describing thermal equilibrium at inverse temperature
$\beta$ and chemical potentials $\mu_1,\ldots,\mu_N$ is given
by the density matrix
\begin{equation}\label{eqrhodef}
\rho_{\beta,\underline{\mu}}:=\Xi_{\beta,\underline{\mu}}^{-1}
\exp[-\beta H_{\underline{\mu}}],
\end{equation}
where
\[H_{\underline{\mu}}:=H-\sum_{i=1}^N\mu_iQ_i,\]
\[\Xi_{\beta,\underline{\mu}}=\mathrm{tr}_{\mathcal H}
[e^{-\beta H_{\underline{\mu}}}];\]
namely
\begin{equation}
\langle a\rangle_{\beta,\underline{\mu}}:=\mathrm{tr}_{\mathcal H}
[\rho_{\beta,\underline{\mu}}a],
\end{equation}
$a\in\mathcal A$. The time-evolution of 
operators in $\mathcal A$ in the Heisenberg
picture is given by
\begin{equation}\label{eqalphadef}
\alpha_t(a):=e^{itH}ae^{-itH}
=e^{itH_{\underline{\mu}}}ae^{-itH_{\underline{\mu}}},
\end{equation}
$a\in\mathcal A$,
where the second equation follows from the fact that elements of 
$\mathcal A$ commute with $Q_1,\ldots,Q_N$. 
From~(\ref{eqrhodef})--(\ref{eqalphadef}) and the cyclicity of the trace
we conclude that
\begin{equation}\label{eqkmsfin}
\langle\alpha_t(a)b\rangle_{\beta,\underline{\mu}}=
\langle b\alpha_{t+i\beta}(a)\rangle_{\beta,\underline{\mu}},
\end{equation}
for arbitrary $a,b$ in $\mathcal A$. This is the famous
{\em KMS condition} characterizing equilibrium states.

\subsection{Systems with Infinitely Many Degrees of Freedom ---
Thermodynamic Limit}
\label{secthermolim}

Systems in non-compact subsets of physical space (e.g.\ the 
thermodynamic limit of physical systems) with infinitely
many degrees of freedom are conveniently described as
{\em$C^*$-dynamical systems}: The algebra of ``observables''
of such a system is thought to be a $C^*$-algebra $\mathcal A$
(with $\|a\|$ the $C^*$-norm of an element $a\in\mathcal A$),
its states are described as {\em normalized, positive linear 
functionals}, $\omega$, on $\mathcal A$;
 (we may assume that
$\mathcal A$ contains an identity element, $\mathbf1$, and that
states are normalized such that $\omega(\mathbf1)=1$).
{\em Symmetries} of such a system are described by a {\em group
of $\star$-automorphisms} of $\mathcal A$. In particular, the
time-translations are described by a one-parameter group,
\begin{equation}
\{\alpha_t(\cdot)|t\in\mathbf R\},
\end{equation}
of $\star$-automorphisms of $\mathcal A$ 
weakly measurable in $t$.

It is convenient to introduce the following subalgebra,
$\Anaught$, of $\mathcal A$:
\begin{equation}\label{eqanaughtdef}
\Anaught:=\left.\left\{a_f\equiv
\int\mathrm dt\,f(t)\alpha_t(a)\right|a\in\mathcal A,
\hat f\in C_0^{\infty}(\mathbf R)\right\},
\end{equation}
where $\hat f$ denotes the Fourier transform of $f$.
Since $\hat f$ is assumed to have compact support, $f(t)$ is
the restriction of an entire function to the real axis.
If the $\star$-automorphisms $\alpha_t$ are norm-continuous 
in $t$, then $\Anaught$ is dense in $\mathcal A$ in norm.
(For definition~(\ref{eqanaughtdef}) to make sense, $\alpha_t(a)$
need only be weakly measurable in $t$. A reasonable hypothesis is to
assume that $\omega(\alpha_t(a))$ is continuous in $t$, for
a weak$^{\star}$-dense set of states, $\omega$, of $\mathcal A$ ---
norm continuity of $\alpha_t$ in $t$ does {\em not} usually
hold.) For an element $a=b_f\in\,\,\Anaught$,
\begin{equation}
\alpha_z(a):=\int\mathrm dt\,f(t-z)\alpha_t(b)
\end{equation}
is {\em entire} in $z$.

As suggested by equation~(\ref{eqkmsfin}) and argued in~[\ref{HHW}],
a state, $\omega_{\beta}$, of such a system describing thermal
equilibrium at inverse temperature $\beta$ should satisfy the 
{\em KMS condition}
\begin{equation}\label{eqkms}
\omega_{\beta}(\alpha_t(a)b)=\omega_{\beta}(b\alpha_{t+i\beta}(a)),
\end{equation}
for all $a\in\Anaught$, $b\in\mathcal A$, $t\in\mathrm R$.
A state $\omega_{\beta}$ satisfying equation~(\ref{eqkms}) is
said to be a ($\beta$-){\em KMS state} for $\alpha_t$. Note that
the KMS condition~(\ref{eqkms}) implies that $\omega_{\beta}$ is
{\em$\alpha_t$-invariant}:
\begin{equation}\label{eqkmsinv}
\omega_{\beta}(\alpha_t(a))=\omega_{\beta}(a),\ a\in\,\,\Anaught.
\end{equation}
(To show~(\ref{eqkmsinv}), one sets $b=\mathbf1$ in~(\ref{eqkms})!)

In order to characterize equilibrium states of infinite systems, the 
KMS condition~(\ref{eqkms}) must be supplemented by an appropriate
{\em``separability-conti\-nuity condition''}. For the purposes of this
paper, the following notion appears to be adequate:

A state $\omega_{\beta}$ of $(\mathcal A,\alpha_t)$ is said to
be an {\em equilibrium state} at inverse temperature $\beta$ iff
\begin{enumerate}
\item \label{omegakms}
$\omega_{\beta}$ is a $\beta$-{\em KMS state} for $\alpha_t$;
\item \label{omegacont}
for arbitrary elements $a$ and $b$ of $\mathcal A$,
\[\omega_{\beta}(a\alpha_t(b))\]
is a {\em continuous} function of $t$;
\item \label{seplty}
the algebra $\mathcal A$ can be given a topology, $\tau$, 
which makes $\mathcal A$ a {\em separable} topological space,
and such that 
$\omega_{\beta}(a\cdot b)$ is {\em jointly continuous
in $a$ and $b$} 
in the product topology on $\mathcal A\times\mathcal A$.
\end{enumerate}

\subsection{The GNS Construction}\label{secgns}

A pair $(\mathcal A,\omega)$ of a $C^*$-algebra $\mathcal A$
and a state $\omega$ of $\mathcal A$ gives rise to a Hilbert
space $\mathcal H_{\omega}$, a $\star$-representation $\lambda_{\omega}$
of $\mathcal A$ on $\mathcal H_{\omega}$, and a cyclic vector 
$\Omega_{\omega}\in\mathcal H_{\omega}$ such that 
\begin{equation}
\omega(a)=\langle\lambda_{\omega}(a)\Omega_{\omega},\Omega_{\omega}\rangle
,\ a\in\mathcal A;
\end{equation}
where $\langle\cdot,\cdot\rangle$ is the scalar product on 
$\mathcal H_{\omega}$; see e.g.~[\ref{L}]. If Property~\ref{seplty},
above,
holds for the state $\omega$, then $\mathcal H_{\omega}$ is {\em separable}.
If $\omega$ is {\em invariant} under a one-parameter $\star$-automorphism
group $\alpha_t$ and $(\mathcal A,\alpha_t,\omega)$ satisfy 
Properties~\ref{omegacont} and \ref{seplty}, above, then there is a 
strongly continuous one-parameter group
\begin{equation}
\left.\left\{e^{itL}\right|t\in\mathbf R\right\}
\end{equation}
of unitary operators, with a self-adjoint generator
\begin{equation}
L=L^*
\end{equation}
such that 
\begin{equation}\label{eqlambdaalpha}
\lambda_{\omega}(\alpha_t(a))=e^{itL}\lambda_{\omega}(a)e^{-itL},
\end{equation}
and 
\begin{equation}\label{eqlambdaomega}
e^{itL}\Omega_{\omega}=\Omega_{\omega},
\end{equation}
for all $t\in\mathbf R$.

We define the kernel, $\mathcal N_{\omega}$, of $\omega$ to be
the left-ideal in $\mathcal A$ given by
\begin{equation}
\mathcal N_{\omega}:=\{a\in\mathcal A|\omega(a^*a)=0\}.
\end{equation}

Let us now assume that $\omega=\omega_{\beta}$ is an
{\em equilibrium state} for $(\mathcal A,\alpha_t)$ at
inverse temperature $\beta$, in the sense that  
Properties~\ref{omegakms}--\ref{seplty} in section~\ref{secthermolim}
hold. Then $\mathcal H_{\beta}:=\mathcal H_{\omega_{\beta}}$ is 
separable (Property~\ref{seplty}), the vector 
$\Omega_{\beta}:=\Omega_{\omega_{\beta}}$ is not only cyclic for
$\lambda(\mathcal A)$, $\lambda\equiv\lambda_{\omega_{\beta}}$, but
{\em separating} (i.e., $\lambda(a)\Omega_{\beta}=0$ implies $\lambda(a)
=0$ on $\mathcal H_{\beta}$, a consequence of the KMS
condition~(\ref{eqkms})),
and $\mathcal N_{\omega_{\beta}}$ is a two-sided $\star$-ideal in 
$\mathcal A$ (Property~\ref{omegakms}, i.e., KMS condition); hence
$\mathcal N_{\omega_{\beta}}=\{0\}$ if $\mathcal A$ is {\em simple}
and $\mbox{dim}\mathcal H_{\beta}>1$.

The generator $L$ is then called {\em thermal Hamiltonian} or
{\em Liouvillian};\linebreak 
(see~[\ref{L},\ref{BR},\ref{HHW}] for further details).

\subsection{Bi-Module Structure of $\mathcal H_{\beta}$ and 
Modular Conjugation~$J$}

The KMS condition gives rise to the following remarkable objects
identified by Haag, Hugenholtz and Winnink in their fundamental
paper~[\ref{HHW}]: It is assumed that $\omega_{\beta}$ is an
equilibrium state for $(\mathcal A,\alpha_t)$ at inverse
temperature $\beta$, in the sense of Properties~\ref{omegakms} 
through \ref{seplty} of section~\ref{secthermolim}.
As noted in section~\ref{secgns}, the vector $\Omega_{\beta}$ 
is then {\em cyclic and separating} for the algebra 
$\lambda(\mathcal A)$. Thus, one can introduce a {\em densely
defined, anti-linear} operator $S$ by
\begin{equation}\label{eqsdef}
S\lambda(a)\Omega_{\beta}:=\lambda(a)^*\Omega_{\beta},\ a\in\mathcal A.
\end{equation}
The KMS condition can be used to show that $S$ can be extended to a
{\em closed} operator and to construct the polar decomposition of
$S$. For this purpose, we define an anti-linear operator $J$ by setting
\[J\lambda(a)\Omega_{\beta}:=
S\lambda(\alpha_{-i\beta/2}(a))\Omega_{\beta}\]
\begin{equation}\label{eqjdef}
=\lambda(\alpha_{i\beta/2}(a^*))\Omega_{\beta},
\end{equation}
for arbitrary $a\in\Anaught$. By Property~\ref{omegacont}, 
$\Omega_{\beta}$ is cyclic and separating for $\lambda(\Anaught)$;
hence $J$ is a densely defined, anti-linear operator. Using the
KMS condition and the invariance of $\omega_{\beta}$ under 
$\alpha_t$, one easily verifies that
\begin{equation}\label{eqjantiunit}
\mbox{$J$ \em is an anti-unitary involution.}
\end{equation}
Using~(\ref{eqsdef})--(\ref{eqjantiunit}) and (\ref{eqlambdaalpha}),
(\ref{eqlambdaomega}), we see that 
\begin{equation}
S=Je^{-\beta L/2}=e^{\beta L/2}J
\end{equation}
is the polar decomposition of $S$.

From (\ref{eqjdef}), (\ref{eqlambdaalpha}) and (\ref{eqlambdaomega}),
\begin{equation}
Je^{itL}=e^{itL}J,\ \forall t\in\mathbf R,
\end{equation}
or, equivalently (recalling that $J$ is {\em anti-linear}),
\begin{equation}
JL=-LJ,
\end{equation}
on the domain of definition of $L$.

One defines
\begin{equation}
\rho(a):=J\lambda(a)J,\ a\in\mathcal A.
\end{equation}
Since $J$ is an anti-unitary involution and $\lambda$ is a 
$\star$-representation of $\mathcal A$, $\rho$ is an
{\em anti-(linear)$\star$-representation} of $\mathcal A$.
By {\em purely algebraic} calculations, one finds that
\begin{equation}\label{eqrholambda}
[\rho(a),\lambda(b)]=0,
\end{equation}
for arbitrary $a,b$ in $\mathcal A$. In fact~[\ref{HHW}],
\begin{equation}\label{eqlambdacomm}
\rho(\mathcal A)''=\lambda(\mathcal A)',
\end{equation}
where $\mathcal B'$ denotes the commutant (commuting algebra)
of an algebra $\mathcal B\subseteq\mathcal L(\mathcal H_{\beta})$,
and $\mathcal B''$ denotes the double commutant.

These results of Haag, Hugenholtz and Winnink contributed to the
development of Tomita-Takesaki theory, see~[\ref{T},\ref{BR}], which is
among the deepest results in the theory of von Neumann algebras.
The starting point is a von Neumann algebra $\mathcal M$ acting
on a Hilbert space $\mathcal H$, with a cyclic and separating
vector $\Omega\in\mathcal H$. One defines
\[SM\Omega=M^*\Omega,\ M\in\mathcal M.\]
It is difficult, but possible, to prove that $S$ can be closed. 
This implies that $\overline S$ has a polar decomposition
\[\overline S=J\exp(-\pi L),\]
where $J$ is an anti-unitary involution, and $L=L^*$ is
self-adjoint. One then proves that
\[\alpha_t(M):=e^{itL}Me^{-itL}\]
is a $\star$-automorphism group of $\mathcal M$, and that
\[\omega(M):=\langle M\Omega,\Omega\rangle\]
is a $2\pi$-KMS state for $(\mathcal M,\alpha_t)$.
As in (\ref{eqrholambda}), (\ref{eqlambdacomm}), it then
follows that
\[\mathcal M':=J\mathcal MJ\]
is the commutant of $\mathcal M$.

The anti-unitarity of $J$ and (\ref{eqrholambda}) may
remind one of the PCT theorem in relativistic QFT and its
proof~[\ref{J2}]. The similarities are not accidental; see
section~\ref{secsscpct}.

Let $(\mathcal A,\alpha_t)$ be a $C^*$-dynamical system,
and $\omega_{\beta}$ an equilibrium state at inverse
temperature $\beta$ for $(\mathcal A,\alpha_t)$, in the 
sense of Properties \ref{omegakms}--\ref{seplty} 
of section~\ref{secthermolim}.
Let us assume that 0 is a {\em simple} eigenvalue of the
Liouvillian $L$ corresponding to the unique eigenvector
$\Omega_{\beta}$. Let $\omega$ be an arbitrary state which
is {\em normal} with respect to $\omega_{\beta}$. Then the
KMS condition for $\omega_{\beta}$ can be used to prove the
property of {\em``return to equilibrium''}; namely
\begin{equation}
\lim_{T\rightarrow\infty}\frac{1}{T}\int_0^T\mathrm dt\,
\omega(\alpha_t(a))=\omega_{\beta}(a),\ a\in\mathcal A,
\end{equation}
which is a remarkable dynamical stability property of KMS
states under local perturbations; see~[\ref{HKT},\ref{JP},\ref{BFS}].

\subsection{Thermal Green Functions and their Analytic
Continuation}

Most properties of a physical system in thermal equilibrium
are encoded in its {\em(real-time) thermal Green functions}
(RTGF), which we define below.

Let $(\mathcal A,\alpha_t)$ be a $C^*$-dynamical system,
and let $\omega_{\beta}$ be an equilibrium state for
$(\mathcal A,\alpha_t)$ at inverse temperature $\beta$, with
Properties \ref{omegakms}--\ref{seplty} 
of section~\ref{secthermolim}. For arbitrary $a_1,\ldots,a_n$
in $\mathcal A$, $t_1,\ldots,t_n$ in $\mathbf R$, we define
\begin{equation}
F_{\beta}(a_1,t_1,\ldots,a_n,t_n):=\omega_{\beta}\left(
\prod_{j=1}^n\alpha_{t_j}(a_j)\right).
\end{equation}
The functions $F_{\beta}$ are the {\em real-time thermal
Green functions}. Because the state $\omega_{\beta}$ is
$\alpha_t$-invariant, they only depend on the variables 
$s_1,s_2,\ldots,s_{n-1}$ defined by
\begin{equation}
t_j=t_1+\sum_{i=1}^{j-1}s_i,\ j=2,\ldots,n.
\end{equation}
If $a_1,\ldots,a_n$ are elements of the algebra $\Anaught$
defined in (\ref{eqanaughtdef}), then
\begin{equation}
H_{\beta}(s_1,\ldots,s_{n-1}):=F_{\beta}(a_1,t_1,\ldots,a_n,t_n)
\end{equation}
is the restriction of an analytic function
$H_{\beta}(\zeta_1,\ldots,\zeta_{n-1})$, 
$(\zeta_1,\ldots,\zeta_{n-1})\in\mathbf C^{n-1}$, to the 
real slice $\mathbf R^{n-1}\subset\mathbf C^{n-1}$. On the
real slice this function is bounded by
\[|H_{\beta}(s_1,\ldots,s_{n-1})|=
|F_{\beta}(a_1,t_1,\ldots,a_n,t_n)|\]
\[=\left|\omega_{\beta}\left(\prod_{j=1}^n\alpha_{t_j}(a_j)\right)\right|\]
\begin{equation}\label{eqrealbound}
\le\prod_{i=1}^n\|a_i\|,
\end{equation}
because $\|\alpha_t(a)\|=\|a\|$, for $a\in\mathcal A$, $t\in\mathbf R$,
and because $\omega_{\beta}$ is a {\em state} on $\mathcal A$. The
KMS condition (\ref{eqkms}) implies that, for every $j=1,\ldots,n-1$,
\pagebreak
\[H_{\beta}(s_1,\ldots,s_j+i\beta,\ldots,s_{n-1})\]
\[=F_{\beta}(a_1,t_1,\ldots,a_j,t_j,a_{j+1},t_{j+1}+i\beta,\ldots,
a_n,t_n+i\beta)\]
\begin{equation}\label{eqhfkms}
=F_{\beta}(a_{j+1},t_{j+1},\ldots,a_n,t_n,a_1,t_1,\ldots,a_j,t_j),
\end{equation}
and thus, as in (\ref{eqrealbound}),
\begin{equation}
|H_{\beta}(s_1,\ldots,s_j+i\beta,\ldots,s_{n-1})|
\le\prod_{i=1}^n\|a_i\|,
\end{equation}
for $j=1,\ldots,n-1$.
By the {\em generalized tube theorem}, due to Kunze, Stein, Malgrange
and Zerner (see e.g.~[\ref{E},\ref{KS}]),
\begin{equation}
|H_{\beta}(\zeta_1,\ldots,\zeta_{n-1})|\le\prod_{i=1}^n\|a_i\|,
\end{equation}
for $a_1,\ldots,a_n$ in $\Anaught$ and $(\zeta_1,\ldots,\zeta_{n-1})$
in the tube
\begin{equation}
T_{n-1}:=\{(\zeta_1,\ldots,\zeta_{n-1})|\mathrm{Im}\zeta_i>0,
\sum_{i=1}^{n-1}\mathrm{Im}\zeta_i<\beta\}.
\end{equation}
It follows that, for $a_1,\ldots,a_n$ in $\Anaught$,
$F_{\beta}(a_1,t_1,\ldots,a_n,t_n)$ is the {\em 
boundary value of a function 
$F_{\beta}(a_1,z_1,\ldots,a_n,z_n)$ analytic
in $(z_1,\ldots,z_n)$ on}
\begin{equation}\label{eqfanal}
\mathcal T_n:=\{(z_1,\ldots,z_n)|\mathrm{Im}z_1<\mathrm{Im}z_2
<\cdots<\mathrm{Im}z_n<\mathrm{Im}z_1+\beta\}
\end{equation}
and bounded on the closure, $\overline{\mathcal T_n}$, of
$\mathcal T_n$ by
\begin{equation}\label{eqfbound}
|F_{\beta}(a_1,z_1,\ldots,a_n,z_n)|\le\prod_{i=1}^n\|a_i\|.
\end{equation}
By Property~\ref{omegacont}, section~\ref{secthermolim}, and
definition~(\ref{eqanaughtdef}), it follows that Properties%
~(\ref{eqfanal}) and (\ref{eqfbound}) hold for {\em arbitrary}
$a_1,\ldots,a_n$ in $\mathcal A$.

These results have first been noticed by Araki~[\ref{A1}].

The functions $F_{\beta}$ have an important {\em positivity
property}. To start  with, we note that, for $a\in\Anaught$,
$z\in\mathbf C$,
\begin{equation}\label{eqalphazstar}
(\alpha_z(a))^*=\alpha_{\overline z}(a^*).
\end{equation}
Let $a_1^i,\ldots,a_{n_i}^i$, $n_i=1,2,\ldots$,  be elements
of $\Anaught$, and $z_1^i,\ldots,z_{n_i}^i$ be complex
numbers with
\begin{equation}\label{eqzinh}
0<\mathrm{Im}z_1^i<\cdots<\mathrm{Im}z_{n_i}^i<\beta/2,
\end{equation}
for $i=1,\ldots,N<\infty$. Let $Z_1,\ldots,Z_N$ be arbitrary
complex numbers and set 
\[a=\sum_{j=1}^NZ_j\prod_{k=1}^{n_j}
\alpha_{z_k^j-i\beta/2}(a_k^j)\in\mathcal A.\]
Since $\omega_{\beta}$ is a state, and by~(\ref{eqalphazstar}),
\[0\le\omega_{\beta}(aa^*)\]
\[=\sum_{i,j=1}^NZ_i\overline Z_jF_{\beta}
(a_1^i,z_1^i-i\beta/2,\ldots,a_{n_i}^i,z_{n_i}^i-i\beta/2,\]
\begin{equation}\label{eqfpos}
(a_{n_j}^j)^*,\overline z_{n_j}^j+i\beta/2,\ldots,
(a_1^j)^*,\overline z_1^j+i\beta/2).
\end{equation}
By invariance, i.e.,
\begin{equation}\label{eqfinv}
F_{\beta}(a_1,z_1+z,\ldots,a_n,z_n+z)=F_{\beta}(a_1,z_1,\ldots,a_n,z_n),
\end{equation}
the positivity~(\ref{eqfpos}) is seen to imply that the complex
numbers
\begin{equation}\label{eqpidef}
\Pi_{ij}:=F_{\beta}(a_1^i,z_1^i,\ldots,a_{n_i}^i,z_{n_i}^i,
(a_{n_j}^j)^*,\overline z_{n_j}^j+i\beta,\ldots,
(a_1^j)^*,\overline z_1^j+i\beta),
\end{equation}
$i,j=1,\ldots,N$, are the matrix elements of a {\em positive
semi-definite matrix $\Pi$}.

Note that, by~(\ref{eqzinh}),
\[(z_1^i,\ldots,z_{n_i}^i,\overline z_{n_j}^j+i\beta,\ldots,
\overline z_1^j+i\beta)\in\mathcal T_{n_i+n_j},\]
for all $i,j$. Thus, by~(\ref{eqfbound})
and Property~\ref{omegacont}, section~\ref{secthermolim},
the positivity Property~(\ref{eqfpos}) holds for {\em arbitrary}
$a_k^i\in\mathcal A$, $z_k^i$ as in~(\ref{eqzinh}), 
$k=1,\ldots,n_i$, $i=1,\ldots,N<\infty$.

We observe  that the KMS condition~(\ref{eqkms}), see also~(\ref{eqhfkms}),
implies that, for arbitrary $a_1,\ldots,a_n$ in $\mathcal A$,
$(z_1,\ldots,z_n)\in\mathcal T_n$,
\begin{equation}\label{eqfkms}
F_{\beta}(a_1,z_1,\ldots,a_n,z_n)
=F_{\beta}(a_{j+1},z_{j+1},\ldots,a_n,z_n,a_1,z_1+i\beta,\ldots,
a_j,z_j+i\beta).
\end{equation}

Finally, Property~\ref{seplty}, section~\ref{secthermolim},
and the KMS condition imply that {\em all} RTGF's and all functions
$F_{\beta}(a_1,z_1,\ldots,a_n,z_n)$ can be obtained as limits of
such functions evaluated on a {\em countable} set of $n$-tuples
$(a_1,\ldots,a_n)$.

Our main results in this section are stated in~(\ref{eqfanal}),
(\ref{eqfbound}) and in~(\ref{eqfinv})--(\ref{eqfkms}).
In particular, (\ref{eqfanal}) shows that we can define 
imaginary-time (``tempe\-rature-ordered'') Green functions
(TOGF's), $\phi_{\beta}$, by setting
\begin{equation}
\phi_{\beta}(a_1,\tau_1,\ldots,a_n,\tau_n):=F_{\beta}(a_1,i\tau_1,
\ldots,a_n,i\tau_n),
\end{equation}
for $a_1,\ldots,a_n$ in $\mathcal A$, and 
\begin{equation}
\tau_1<\tau_2<\cdots<\tau_n<\tau_1+\beta.
\end{equation}
It is convenient to think of $\tau_1,\ldots,\tau_n$ as 
{\em angles} on a circle of circumference $\beta$, ordered
in accordance with the orientation chosen on the circle;
see~(\ref{eqfinv}) and (\ref{eqfkms}).

The main properties of TOGF's can immediately be inferred from~%
(\ref{eqfbound}) and (\ref{eqfinv})--(\ref{eqfkms}). In the next
section, we show that functions with all the general properties
of TOGF's {\em are}, in fact,  the TOGF's corresponding
to an equilibrium state, $\omega_{\beta}$, of a $C^*$-dynamical
system.

\section{A Reconstruction Theorem at Positive Temperature}
\label{secrecn}
\input{sec3}

\section{KMS $\leftrightarrow$ SSC, PCT, (A)dS, etc.}
\label{secqft}
\input{sec4}

\newpage
\textbf{\Large Bibliography}\newline
\begin{enumerate}

\item\label{G}
J. Ginibre, {\em Reduced Density Matrices of Quantum Gases},
J. Math.\ Phys.~\textbf{6}, 238--251, 
252--262, 1432--1446; and
in {\em Statistical Mechanics}, ed.\ T. Bak, Benjamin, New York,
1967, p.\ 148

\item\label{FKT}
J. Feldman and E. Trubowitz, {\em Perturbation Theory for
Many Fermion Systems},
Helv.\ Phys.\ Acta \textbf{63},
156--260 (1990),
J. Feldman, H. Kn\"orrer and E. Trubowitz,
{\em A Two-Dimensional Fermi Liquid}, {\em Single Scale
Analysis of Many-Fermion Systems}, {\em Convergence of
Perturbation Expansions in Fermionic Models}, \&c.,
see \texttt{http://www.math.ubc.ca/people}
\texttt{/faculty/feldman/fl.html} 

\item\label{CFS}
T. Chen, J. Fr\"ohlich and M. Seifert, {\em Renormalization Group
Methods: Landau-Fermi Liquid and BCS Superconductor},
in: {\em Fluctuating Geometries in Statistical Mechanics and
Field Theory}, F. David, P. Ginsparg and J. Zinn-Justin, eds.,
Les Houches Summer School 1994, Elsevier Science, Amsterdam, 1995

\item\label{OS}
K. Osterwalder and R. Schrader, {\em Axioms for Euclidean Green's
Functions}, Commun.\ Math.\ Phys.\ \textbf{42}, 281--305 (1975)

\item\label{Gl}
V. Glaser, {\em On the Equivalence of the Euclidean and Wightman
Formulation of Field Theory}, Comm.\ Math.\ Phys.\ \textbf{37},
257--272 (1974)

\item\label{J}
R. Jost, {\em The General Theory of Quantized Fields}, American 
Mathematical Society, Providence, Rhode Island, 1965

\item\label{StW}
R.F. Streater and A.S. Wightman, {\em PCT, Spin and Statistics,
and All That}, Benjamin, New York, 1964

\item\label{K} 
R. Kubo, J. Phys.\ Soc.\ Japan \textbf{12}, 570 (1957)

\item\label{MS}
P.C. Martin and J. Schwinger, Phys. Rev. \textbf{115}, 1342 (1959)

\item\label{Ru}
D. Ruelle, {\em Analyticity of Green's Functions of Dilute
Quantum Gases}, J. Math.\ Phys.\ \textbf{12}, 901--903,
(1971), and {\em Definition of Green's Functions for Dilute 
Fermi Gases}, Helv.\ Phys.\ Acta \textbf{45}, 215--219 (1972)

\item\label{HHW}
R. Haag, N. Hugenholtz and M. Winnink, {\em On the Equilibrium
States in Quantum Statistical Mechanics}, Commun.\ Math.\ Phys.\ 
\textbf{5}, 215--236 (1967)

\item\label{A1}
H. Araki, {\em Multiple Time Analyticity of a Quantum
Statistical State Satisfying the KMS Boundary Condition},
Publ.\ RIMS, Kyoto Univ.\ Ser.\ A, \textbf{4}, 361--371 (1968)

\item\label{HK}
R. H\o gh-Krohn, {\em Relativistic Quantum Statistical Mechanics
in Two-Dimensional Space-Time}, Comm.\ Math.\ Phys.\ \textbf{38},
195--224 (1974),
R. Figari, R. H\o gh-Krohn and C.R. Nappi, {\em Interacting
Relativistic Boson Fields in the De Sitter Universe with Two
Space-Time Dimensions}, Comm.\ Math.\ Phys.\ \textbf{44},
265--278 (1975)

\item\label{F1}
J. Fr\"ohlich, {\em The Reconstruction of Quantum Fields
from Euclidean Green's Functions at Arbitrary Temperatures},
Helv.\ Phys.\ Acta \textbf{48}, 355--369 (1975)

\item\label{F2}
J. Fr\"ohlich, Lectures at Princeton University 1976/77, 
unpublished

\item\label{A2}
H. Araki, {\em Relative Hamiltonian for
Faithful Normal States of a von Neumann Algebra}, Publ.\ RIMS Kyoto Univ.,
\textbf{9}, 165--209 (1973)

\item\label{F3}
J. Fr\"ohlich, {\em Unbounded Symmetric Semigroups on a Separable
Hilbert Space Are Essentially Selfadjoint}, Adv.\ Appl.\ Math.\
\textbf{1}, 237--256 (1980)

\item\label{KL1}
A. Klein and L.J. Landau, {\em Construction of a Unique Self-Adjoint
Generator for a Symmetric Local Semigroup}, J. Funct.\ Anal.\ 
\textbf{44}, 121--137 (1981)

\item\label{KL2}
A. Klein and L.J. Landau, {\em Stochastic Processes Associated with
KMS States}, J.\ Funct.\ Anal.\ \textbf{42}, 368--428 (1981)

\item\label{BW}
J.J. Bisognano and E.H. Wichmann, {\em On the Duality Condition for
a Hermitian Scalar Field}, J. Math.\ Phys.\ \textbf{16}, 985--1007
(1975)

\item\label{J2}
R. Jost, {\em Eine Bemerkung zum CTP Theorem}, Helv.\ Phys.\ Acta
\textbf{30}, 409--416 (1957)

\item\label{FOS}
J. Fr\"ohlich, K. Osterwalder and E. Seiler, {\em On Virtual Representations
of Symmetric Spaces and their Analytic Continuation}, Ann.\ Math.\ 
\textbf{118}, 461--489 (1983)

\item\label{L} 
O.E. Lanford, in: {\em M\'ecanique Statistique et Th\'eorie
Quantique des Champs}, C. De Witt, R. Stora, eds., Gordon and
Breach, New York, 1971 (Les Houches Summer School 1970), 
pp.\ 109--214

\item\label{BR}
O. Bratteli and D.W. Robinson, {\em Operator Algebras and Quantum
Statistical Mechanics}, Springer Verlag, New York, Volume I: 1979,
Volume II: 1981

\item\label{T}
M. Takesaki, {\em Tomita's Theory of Modular Hilbert Algebras
and its Applications}, LNM \textbf{128}, Springer Verlag, Berlin,
Heidelberg, 1970

\item\label{HKT} 
R. Haag, D. Kastler and E.B. Trych-Pohlmeyer, {\em Stability
and Equilibrium States}, Comm.\ Math.\ Phys.\ \textbf{38},
173--193 (1974)

\item\label{JP} 
V. Jak\v si\'c and C.A. Pillet, {\em On a Model for Quantum
Friction III. Ergodic Properties of the Spin-Boson System}, 
Comm.\ Math.\ Phys.\ 
\textbf{178}, 627--651 (1996)

\item\label{BFS}
V. Bach, J. Fr\"ohlich and I.M. Sigal, {\em Return to Equilibrium},
J. Math.\ Phys.\ \textbf{41}, 3985 (2000)

\item\label{E}
H. Epstein, in: {\em Axiomatic Field Theory}, M. Chretien and S. Deser
eds., Gordon and Breach, New York, 1966 (Brandeis Summer School 1965)

\item\label{KS}
H. Kunze and E. Stein, {\em Uniformly Bounded Representations II},
Amer.\ J. Math.\ \textbf{83}, 723--786 (1960)

\item\label{KL3}
A. Klein and L.J. Landau, {\em From the Euclidean Group to the
Poincar\'e Group via Osterwalder-Schrader Positivity}, Comm.\ Math.\
Phys.\ \textbf{87}, 469--484 (1983)

\item\label{FM}
J. Fr\"ohlich and P.A. Marchetti, {\em Spin-Statistics Theorem and 
Scattering in Two-Dimensional Condensed Matter Physics}, Nucl.\
Phys.\ \textbf{B356}, 533--573 (1991)

\item\label{S}
G.L. Sewell, {\em Relativity of Temperature and the Hawking Effect},
Phys.\ Lett.\ \textbf{79A}, 23/24 (1980)

\item\label{BEM}
J. Bros, H. Epstein and U. Moschella, {\em Analyticity Properties 
and Thermal Effects for General Quantum Field Theory on De Sitter
Space-Time}, Comm.\ Math.\ Phys.\ \textbf{196}, 535--570 (1998)

\item\label{LM}
M. L\"uscher and G. Mack, {\em Global Conformal Invariance in
Quantum Field Theory}, Comm.\ Math.\ Phys.\ \textbf{41},
203--234 (1975)

\item\label{Fr}
J. Fr\"ohlich, {\em Statistics of Fields, the Yang-Baxter Equation
and the Theory of Knots and Links},
in: G. 't Hooft et al., eds., 
{\em Non-Perturbative Quantum Field Theory}, 
Carg\`ese 1987,
NATO Advanced Science Institutes Series B: Physics
\textbf{185}, 71--100, Plenum, New York

\item\label{A3}
H. Araki, {\em On the Connection of Spin and Commutation Relations
between Different Fields}, J. Math.\ Phys.\ \textbf{2}, 267--270 
(1961)

\item\label{FG}
J. Fr\"ohlich and F. Gabbiani, 
{\em Braid Statistics in Local Quantum Theory}, Rev.\ Math.\ Phys.\
\textbf{2}, 251--353 (1990)

\item\label{BM}
J. Bros, H. Epstein and U. Moschella, {\em Towards a General Theory
of Quantized Fields on the  Anti-de Sitter Space-Time}, 
preprint November 2001

\item\label{Mal}
J. Maldacena, {\em The Large $N$ Limit of Superconformal Field
Theories and Supergravity}, Adv.\ Theor.\ Math.\ Phys.\ \textbf{2},
231--252 (1998)

\item\label{St}
N. Straumann, {\em General Relativity and Relativistic Astrophysics},
Springer-Verlag, Berlin, Heidelberg, New York, 1984

\end{enumerate}

\end{document}

%% file: sec3.tex
In this section, we show how to reconstruct the RTGF's of a
$C^*$-dynamical system in an equilibrium state from
functions with all the properties of TOGF's. Our result is an
analogue of the {\em Osterwalder-Schrader reconstruction
theorem} [\ref{OS},\ref{Gl}], which has solved a similar problem
at zero temperature. A result of the kind we shall prove in this
section, but with additional assumptions that make it inapplicable
to systems of {\em fermions}, for example,
such as non-relativistic electron
liquids (see~[\ref{FKT}]), has been proven in~[\ref{KL2}]; see
[\ref{Ru},\ref{HK},\ref{F1},\ref{F2},\ref{F3}] for 
earlier, partial results.

\subsection{Green Functions on an (Imaginary-Time) Circle}
\label{secfprop}

Our starting point, in this section, is a set of {\em Green
functions} depending on $n$-tuples $[a_1,\tau_1,\ldots,a_n,\tau_n]$,
where $a_i$ is an element of a {\em separable topological space
$\mathcal S$}, $\tau_i$ is a point on a circle of circumference
$\beta$, for all $i=1,\ldots,n$, and 
$(\tau_1,\ldots,\tau_n)\in\mathbf T^n_<$, where
\begin{equation}\label{eqtdef}
\mathbf T^n_<:=\{(\sigma_1,\ldots,\sigma_n)|\sigma_1<\sigma_2<\cdots
<\sigma_n<\sigma_1+\beta\}.
\end{equation}
These Green functions are denoted
\begin{equation}
\phi_{\beta}(a_1,\tau_1,\ldots,a_n,\tau_n),
\end{equation}
$n=0,1,2,\ldots$, with $\phi_{\beta}(\emptyset)=1$.
They are assumed to have the following properties.
For arbitrary $a_1,\ldots,a_n$ in $\mathcal S$, and
$n=0,1,2,\ldots$:

\begin{description}
\item[(P1) Continuity:]
$\phi_{\beta}(a_1,\tau_1,\ldots,a_n,\tau_n)$ is defined
for arbitrary $(a_1,\ldots,a_n)$ $\in\mathcal S^{\times n}$
and $(\tau_1,\ldots,\tau_n)\in\mathbf T^n_<$; it is jointly
continuous in $(a_1,\ldots,a_n)$ in the product topology
of $\mathcal S^{\times n}$, and it is a continuous 
function of $(\tau_1,\ldots,\tau_n)$ on $\mathbf T^n_<$.

\item[(P2) Translation invariance:]
\[\phi_{\beta}(a_1,\tau_1,\ldots,a_n,\tau_n)
=\phi_{\beta}(a_1,\tau_1+\tau,\ldots,a_n,\tau_n+\tau),\]
for arbitrary $\tau\in\mathbf R$.

\item[(P3) KMS condition:]
\[\phi_{\beta}(a_1,\tau_1,\ldots,a_n,\tau_n)
=\phi_{\beta}(a_{j+1},\tau_{j+1},\ldots,a_n,\tau_n,
a_1,\tau_1+\beta,\ldots,a_j,\tau_j+\beta),\]
for arbitrary $j=1,\ldots,n-1$.

\item[(P4) Reflection positivity:]
There is a continuous involution $^*$ on $\mathcal S$,
\[\mathcal S\ni a\mapsto a^*\in\mathcal S,\ 
(a^*)^*=a,\ \forall a,\]
with the property that, for all $N=1,2,3,\ldots$,
arbitrary $a_1^i,\ldots,a_{n_i}^i$ in $\mathcal S$,
$n_i=0,1,2,\ldots$, $i=1,\ldots,N$, the matrix
$\Pi=(\Pi_{ij})_{i,j=1,\ldots,N}$, defined by
\begin{equation}\label{eqbigpidef}
\Pi_{ij}:=\phi_{\beta}(a_1^i,\tau_1^i,\ldots,a_{n_i}^i,\tau_{n_i}^i,
(a_{n_j}^j)^*,\beta-\tau_{n_j}^j,\ldots,(a_1^j)^*,\beta-\tau_1^j),
\end{equation}
with
\begin{equation}\label{eqtaurangenoncoin}
0<\tau_1^i<\cdots<\tau_{n_i}^i<\beta/2,\ \forall i,
\end{equation}
is {\em positive semi-definite}.
\end{description}

In much of this section, we shall require a much stronger version
of Property~(P1), namely:
\begin{description}
\item[(P$^*$) TOGF's on a $C^*$-algebra:]
The space $\mathcal S$ is a $C^*$-algebra with identity,
$\mathbf1$, and the involution $^*$ in (P4) is the usual
$\star$-operation on $\mathcal S$. It is then assumed that

\item[(P$^*$i)]
$\phi_{\beta}(a_1,\tau_1,\ldots,a_n,\tau_n)$ is {\em linear}
in each argument $a_i$, $i=1,\ldots,n$, {\em jointly
continuous} in $(a_1,\ldots,a_n)$ in the product topology
on $\mathcal S^{\times n}$ of a topology on $\mathcal S$
in which $\mathcal S$ is separable, and {\em continuous}
in $\tau_1,\ldots,\tau_n$ on the closure, 
$\overline{\mathbf T^n_<}$,
of $\mathbf T^n_<$;

\item[(P$^*$ii)]
\[\phi_{\beta}(a_1,\tau_1,\ldots,a_j,\tau_j,a_{j+1},\tau_j,
\ldots, a_n,\tau_n)\]\[=
\phi_{\beta}(a_1,\tau_1,\ldots,a_j\cdot a_{j+1},\tau_j,
\ldots, a_n,\tau_n),\]
for arbitrary $j=1,\ldots,n-1$, $n=2,3,\ldots$;

\item[(P$^*$iii)]
\[\phi_{\beta}(a_1,\tau_1,\ldots,a_{j-1},\tau_{j-1},
\mathbf1,\tau_j,\ldots,a_n,\tau_n)\]\[=
\phi_{\beta}(a_1,\tau_1,\ldots,a_{j-1},\tau_{j-1},
a_{j+1},\tau_{j+1},\ldots,a_n,\tau_n),\]
for arbitrary $j=1,\ldots,n$; and

\item[(P$^*$iv)]
\[|\phi_{\beta}(a_1,\tau_1,\ldots,a_n,\tau_n)|\le
\prod_{j=1}^n\|a_j\|,\]
where $\|(\cdot)\|$ is the $C^*$-norm on $\mathcal S$.
\end{description}

\textbf{Remark:} In the last section, we have seen that 
Properties~(P1)--(P4) and~(P$^*$) hold for the TOGF's
associated with an equilibrium state, $\omega_{\beta}$,
of a $C^*$-dynamical system $(\mathcal A,\alpha_t)$,
with $\mathcal S=\mathcal A$.

It may be appropriate to mention some examples of physical systems
with TOGF's satisfying Properties~(P1)--(P4) and~(P$^*$):

(1) Let $\mathcal S$ be the CAR algebra of a system of non-relativistic
fermions of the kind considered by Ginibre in~[\ref{G}], and let
$\phi_{\beta}(a_1,\tau_1,\ldots,a_n,\tau_n)$ be the TOGF's of such
a system as constructed in~[\ref{G},\ref{Ru}],
for sufficiently small $\beta$. The functional-integral definition
of $\phi_{\beta}(a_1,\tau_1,\ldots,a_n,\tau_n)$ makes it clear that
these functions can be defined for arbitrary $n$-tuples
$(\tau_1,\ldots,\tau_n)$, and if $a_1,\ldots,a_n$ are 
creation- or annihilation operators then 
$\phi_{\beta}(a_1,\tau_1,\ldots,a_n,\tau_n)$ is totally 
antisymmetric in its $n$ arguments $(a_i,\tau_i)$,
$i=1,\ldots,n$. If $\Psi^*$ and $\Psi$ denote a creation-
and the corresponding annihilation operator in $\mathcal S$, then
\pagebreak
\[\phi_{\beta}(\Psi^*,\tau_1,\Psi,\tau_2)\stackrel{\mbox{KMS}}{=}
\phi_{\beta}(\Psi,\tau_2,\Psi^*,\tau_1+\beta)\]
\begin{equation}
=-\phi_{\beta}(\Psi^*,\tau_1+\beta,\Psi,\tau_2).
\end{equation}
Thus, $\phi_{\beta}(\Psi^*,\tau_1,\Psi,\tau_2)$ is an 
{\em anti-periodic} function of $\tau_1-\tau_2\in[0,\beta]$.

(2) For systems of non-relativistic bosons or of Bose quantum
fields, as considered in~[\ref{HK},\ref{F1},\ref{KL2}], one
may choose $\mathcal S$ to be a $C^*$-algebra generated
by {\em Weyl operators} constructed from bosonic creation-
and annihilation operators. For {\em bosons}, the creation-
and annihilation operators, $\Phi^*$, $\Phi$, are
{\em unbounded} operators (in contrast to the {\em bounded}
creation- and annihilation operators for {\em fermions}).
Yet, it may happen that, for arbitrary $n$, the TOGF's
\[\phi_{\beta}(\Phi_1^{\#},\tau_1,\ldots,\Phi_n^{\#},\tau_n)\]
are well defined; here $\Phi_j^{\#}=\Phi_j$ or $\Phi_j^*$,
for all $j$.  The TOGF's turn
out to be totally symmetric under permutations of their
arguments. Hence, the KMS condition implies that
\[\phi_{\beta}(\Phi^*,\tau_1,\Phi,\tau_2)\stackrel{\mbox{KMS}}{=}
\phi_{\beta}(\Phi,\tau_2,\Phi^*,\tau_1+\beta)\]
\begin{equation}
=\phi_{\beta}(\Phi^*, \tau_1+\beta,\Phi, \tau_2),
\end{equation}
i.e.\ $\phi_{\beta}(\Phi^*,\tau_1,\Phi,\tau_2)$ is a {\em periodic}
function of $\tau_1-\tau_2\in[0,\beta]$.

\subsection{The Main Theorem}
\label{secmainthm}

In this section, we describe our main result concerning the
reconstruction of a thermal equilibrium state and of 
real-time Green functions from a set of TOGF's  with
the properties of section~\ref{secfprop}. Let $\mathcal  S$,
$\mathbf T^n_<$, etc.\ be as in section~\ref{secfprop}.

\medskip\medskip

\textbf{Main Theorem}

{\em
(1) Assume that the TOGF's
\[\{\phi_{\beta}(a_1,\tau_1,\ldots,a_n,\tau_n)\}_{n=0}^{\infty}\]
have Properties~(P1)--(P4) of section~\ref{secfprop}. Then they
uniquely determine a separable Hilbert space $\mathcal H_{\beta}$,
a continuous, unitary one-parameter group $\{e^{itL}\}_{t\in\mathbf R}$
on $\mathcal H_{\beta}$, a vector $\Omega_{\beta}\in\mathcal H_{\beta}$
invariant under $\{e^{itL}\}_{t\in\mathbf R}$, and an anti-unitary
operator $J$ on $\mathcal H_{\beta}$ such that 
\begin{equation}
J\Omega_{\beta}=\Omega_{\beta},\ e^{itL}J=Je^{itL}.
\end{equation}

(2) If $\mathcal S$ is a $C^*$-algebra, and, in addition to~%
(P1)--(P4), Property (P$^*$) of section~\ref{secfprop} holds, then
the TOGF's
\[\{\phi_{\beta}(a_1,\tau_1,\ldots,a_n,\tau_n)\}_{n=0}^{\infty}\]
determine a $\star$-representation, $\lambda$, of $\mathcal S$
on $\mathcal H_{\beta}$ and an anti-representation, $\rho$,
of $\mathcal S$ on $\mathcal H_{\beta}$ given by
\begin{equation}
\rho(a)=J\lambda(a)J,\ a\in\mathcal S,
\end{equation}
such that
\begin{equation}
\left[e^{itL}\lambda(a)e^{-itL},e^{isL}\rho(b)e^{-isL}\right]=0,
\end{equation}
for all $a,b$ in $\mathcal S$ and $t,s$ real.
The state
\begin{equation}
\omega_{\beta}(\cdot):=\langle(\cdot)\Omega_{\beta},\Omega_{\beta}\rangle
\end{equation}
is a KMS state for $\lambda(\mathcal S)$ and the time evolution
\[\lambda(a)\mapsto e^{itL}\lambda(a)e^{-itL},\ a\in\mathcal S.\]
The functions $\{\phi_{\beta}(a_1,\tau_1,\ldots,a_n,\tau_n)\}_{n=0}^{\infty}$
are the TOGF's obtained from the real-time Green functions
\begin{equation}\label{eqgreenfunc}
\left\langle\prod_{j=1}^ne^{it_jL}\lambda(a_j)e^{-it_jL}\Omega_{\beta},
\Omega_{\beta}\right\rangle
\end{equation}
by analytic continuation in the time variables $t_1,\ldots,t_n$ to the
tube $\mathcal T_n$ defined in~(\ref{eqfanal}) and restriction to
\[\{t_j=i\tau_j|j=1,\ldots,n,\ \tau_1<\tau_2<\cdots<\tau_n<\tau_1+\beta\}.\]

}

\textbf{Remarks:} A similar result, but in a more special situation, has
been established by Klein and Landau in~[\ref{KL2}];
 (the results
in~[\ref{KL2}] do not apply to systems of fermions, for example). 
With the exception
of the very last part, this theorem was proven in~[\ref{F2}]; see also~%
[\ref{FOS},\ref{KL3}] for further results.

Our result is an analogue, at positive temperature, of the
Osterwalder-Schrader reconstruction theorem~[\ref{OS},\ref{Gl}].

The proof of the Main Theorem forms the core of our paper.

\subsection{Proof of Part (1) of the Main Theorem}
\label{secproofpart1}

The proof of the Main Theorem consists of a highly non-trivial
extension of the GNS construction. The first step is to
construct the Hilbert space $\mathcal H_{\beta}$.

\subsubsection{i) Construction of Hilbert space}

We consider the linear space
\begin{equation}\label{eqvdef}
\mathcal V_{\beta}:=\bigoplus_{n=0}^{\infty}\mathcal V_{\beta}^{(n)},
\end{equation}
of formal polynomials, where
\begin{equation}
\mathcal V_{\beta}^{(n)}:=\left\{\sum_iZ_i[a_1^i,\tau_1^i,\ldots,
a_n^i,\tau_n^i]\right\},
\end{equation}
with $Z_i\in\mathbf C$, $a_1^i,\ldots,a_n^i$ in $\mathcal S$,
\begin{equation}
0<\tau_1^i<\cdots<\tau_n^i<\beta/2,
\end{equation}
for all $i$, and
\begin{equation}
\mathcal V_{\beta}^{(0)}:=\mathbf C.
\end{equation}
The space $\mathcal V_{\beta}$ can be equipped with a positive
semi-definite inner product determined from
\[\left\langle[a_1,\tau_1,\ldots,a_n,\tau_n],[b_1,\sigma_1,\ldots,b_m,\sigma_m]
\right\rangle:=\]
\begin{equation}\label{eqproddef}
\phi_{\beta}(a_1,\tau_1,\ldots,a_n,\tau_n,b_m^*,
\beta-\sigma_m,\ldots,b_1^*,\beta-\sigma_1),
\end{equation}
by linearity in the first and anti-linearity in the second argument;
$0<\tau_1<\cdots<\tau_n<\beta/2$, $0<\sigma_1<\cdots<\sigma_m<\beta/2$.
The reflection positivity property, (P4), implies that, indeed,
(\ref{eqproddef}) determines a positive (semi-)definite inner
product on $\mathcal V_{\beta}$. We define the kernel of 
$\langle\cdot,\cdot\rangle$ by
\begin{equation}\label{eqndef}
\mathcal N_{\beta}:=\{v\in\mathcal V_{\beta}|\langle v,v\rangle=0\}.
\end{equation}

The equivalence class modulo $\mathcal N_{\beta}$ of an element
$v\in\mathcal V_{\beta}$ is denoted by
\begin{equation}\label{eqphiv}
\Phi(v):=v\bmod\mathcal N_{\beta}.
\end{equation}

Clearly,
\begin{equation}
\mathcal H_{\beta}:=\overline{\mathcal V_{\beta}/\mathcal N_{\beta}},
\end{equation}
where the closure is taken in the norm determined by the scalar
product $\langle\cdot,\cdot\rangle$ on
$\mathcal V_{\beta}/\mathcal N_{\beta}$, is a Hilbert space.

By Property~(P1) and the separability of $\mathcal S$, $\mathcal H_{\beta}$
is a {\em separable} Hilbert space.

By construction, the linear space
\begin{equation}
\mathcal D_{\beta}:=\Phi(\mathcal V_{\beta})
\end{equation}
is {\em dense} in $\mathcal H_{\beta}$. We define the vector 
$\Omega_{\beta}$ by
\begin{equation}
\Omega_{\beta}=\Phi([\emptyset]),
\end{equation}
with
\[\langle\Omega_{\beta},\Omega_{\beta}\rangle=\phi_{\beta}
(\emptyset):=1.\]

\subsubsection{ii) Construction of a unitary one-parameter group of
time translations}

By linearity, the equation
\begin{equation}\label{eqtaushift}
[a_1,\tau_1,\ldots,a_n,\tau_n]_{\tau}:=
[a_1,\tau_1+\tau,\ldots,a_n,\tau_n+\tau],
\end{equation}
for $-\tau_1<\tau<\beta/2-\tau_n$ ($0<\tau_1<\cdots<\tau_n<\beta/2$,
$a_1,\ldots,a_n$ in $\mathcal S$), defines a {\em shift operator}
\begin{equation}
\mathcal V_{\beta}\ni v\mapsto v_{\tau}\in\mathcal V_{\beta},
\end{equation}
for all $\tau\in(-\epsilon_-(v),\epsilon_+(v))$, for some
positive numbers $\epsilon_-(v)$ and $\epsilon_+(v)$ (with
$\epsilon_-(v)=\tau_1$, $\epsilon_+(v)=\beta/2-\tau_n$, for $v$
as in~(\ref{eqtaushift})).

It is clear from~(\ref{eqtaushift}) that
\begin{equation}
(v_{\tau})_{\sigma}=v_{\tau+\sigma},
\end{equation}
if $\tau$, $\sigma$ and $\tau+\sigma$ all belong to the open
interval $(-\epsilon_-(v),\epsilon_+(v))$.

Let $v$ and $w$ be two vectors in $\mathcal V_{\beta}$.
Then the definition~(\ref{eqproddef}) of the inner product
and Property~(P2) (translation invariance) readily imply that
\begin{equation}\label{eqshiftsym}
\langle v_{\tau},w\rangle=\langle v,w_{\tau}\rangle
\end{equation}
if $-\min(\epsilon_-(v),\epsilon_-(w))<\tau<\min(\epsilon_+(v),
\epsilon_+(w))$.

We claim that
\begin{equation}\label{eqninv}
v\in\mathcal N_{\beta}\Rightarrow v_{\tau}\in\mathcal N_{\beta},
\mbox{ for }-\epsilon_-(v)<\tau<\epsilon_+(v).
\end{equation}
To prove~(\ref{eqninv}), we notice that, for 
$-\epsilon_-(v)/2<\tau<\epsilon_+(v)/2$,
\[0\le\langle v_{\tau},v_{\tau}\rangle=
\langle v,v_{2\tau}\rangle\]
\[\le\langle v,v\rangle^{1/2}\langle v_{2\tau},v_{2\tau}\rangle^{1/2}\]
\begin{equation}\label{eqninvind}
=0,
\end{equation}
 for $v\in\mathcal N_{\beta}$,
by the Cauchy-Schwarz inequality; hence $v_{\tau}\in\mathcal N_{\beta}$.
For $\tau\in(-\epsilon_-(v)/2,\epsilon_+(v)/2)$ and
$\tau_1\in(-\epsilon_-(v)/4,\epsilon_+(v)/4)$, we have that 
$v_{\tau+\tau_1}$ and $v_{\tau+2\tau_1}$ are in $\mathcal V_{\beta}$,
and
\[0\le\langle v_{\tau+\tau_1},v_{\tau+\tau_1}\rangle=
\langle v_{\tau},v_{\tau+2\tau_1}\rangle\]
\[\le\langle v_{\tau},v_{\tau}\rangle^{1/2}
\langle v_{\tau+2\tau_1},v_{\tau+2\tau_1}\rangle^{1/2}\]
\[=0,\]
because $v_{\tau}\in\mathcal N_{\beta}$, by~(\ref{eqninvind}).
This makes it clear that the proof of~(\ref{eqninv}) can be
completed inductively.

Observation~(\ref{eqninv}) permits us to define operators,
$\Gamma_{\tau}$, on the dense domain 
$\mathcal D_{\beta}\subset\mathcal H_{\beta}$ as follows:
Each $\Psi\in\mathcal D_{\beta}$ is of the form
$\Psi=\Phi(v)$, for some $v\in\mathcal V_{\beta}$. For
$\tau\in(-\epsilon_-(v),\epsilon_+(v))$, we set
\begin{equation}\label{eqgammadef}
\Gamma_{\tau}\Psi:=\Phi(v_{\tau}).
\end{equation}
Defining
\begin{equation}
\epsilon_{\pm}(\Psi):=\sup_{v\in\mathcal V_{\beta}}
\{\epsilon_{\pm}(v)|\Phi(v)=\Psi\},
\end{equation}
we see that~(\ref{eqninv}) implies that the left hand side
of~(\ref{eqgammadef}) is well defined, for 
$\tau\in(-\epsilon_-(\Psi),\epsilon_+(\Psi))$. Property~(P1)
(continuity) then implies that
\begin{equation}
\mbox{s-}\!\!\lim_{\tau\rightarrow0}\Gamma_{\tau}\Psi=\Psi,\ 
\forall\Psi\in\mathcal D_{\beta}.
\end{equation}
Next, for $\Psi=\Phi(v)$ and $\tilde{\Psi}=\Phi(\tilde v)$
in $\mathcal D_{\beta}$, and
$-\min(\epsilon_-(\Psi),\epsilon_-(\tilde{\Psi}))<\tau<
\min(\epsilon_+(\Psi),\epsilon_+(\tilde{\Psi}))$,
\[\langle\Gamma_{\tau}\Psi,\tilde{\Psi}\rangle=
\langle\Gamma_{\tau}\Phi(v),\Phi(\tilde v)\rangle\]
\[\hspace{6 cm}=\langle v_{\tau},\tilde v\rangle,
\hspace{3 cm}\mbox{ by (\ref{eqgammadef}),(\ref{eqphiv})}\]
\[\hspace{6 cm}=\langle v,\tilde v_{\tau}\rangle,
\hspace{4 cm}\mbox{ by (\ref{eqshiftsym})}\]
\[=\langle\Phi(v),\Gamma_{\tau}\Phi(\tilde v)\rangle\]
\begin{equation}
=\langle\Psi,\Gamma_{\tau}\tilde{\Psi}\rangle.
\end{equation}
Finally, for $\Psi=\Phi(v)\in\mathcal D_{\beta}$, and 
$\tau,\sigma,\tau+\sigma$ all in the interval 
$(-\epsilon_-(\Psi),\epsilon_+(\Psi))$,
\[\Gamma_{\tau}(\Gamma_{\sigma}\Psi)=
\Gamma_{\tau}(\Gamma_{\sigma}\Phi(v))\]
\[=\Gamma_{\tau}\Phi(v_{\sigma})\]
\[=\Phi(v_{\tau+\sigma})\]
\[=\Gamma_{\tau+\sigma}\Phi(v)\]
\begin{equation}\label{eqgammagrp}
=\Gamma_{\tau+\sigma}\Psi.
\end{equation}
A somewhat remarkable {\em theorem on the essential self-adjointness
of local, Hermitian semigroups} proven in~[\ref{F3},\ref{KL1}]
says that from (\ref{eqgammadef}) through~(\ref{eqgammagrp})
it follows that
\begin{equation}\label{eqgammal}
\Gamma_{\tau}\Psi=e^{\tau L}\Psi,\mbox{ for }
-\epsilon_-(\Psi)<\tau<\epsilon_+(\Psi),
\end{equation}
{\em for every $\Psi\in\mathcal D_{\beta}$, where $L$, the
``Liouvillian'', is essentially self-adjoint on a domain
$\stackrel{\,\,\,\circ}{\mathcal D}_{\beta}\subset
\mathcal D_{\beta}$ which is dense in $\mathcal H_{\beta}$.}
(In~[\ref{F3}], there is an explicit construction of 
$\stackrel{\,\,\,\circ}{\mathcal D}_{\beta}$.)

Clearly,
\[\Gamma_{\tau}\Omega_{\beta}=\Gamma_{\tau}\Phi([\emptyset])
=\Phi([\emptyset])=\Omega_{\beta},\]
for arbitrary $\tau$, i.e.,
\begin{equation}
L\Omega_{\beta}=0.
\end{equation}
By Stone's theorem, $\exp(itL)|_{t\in\mathbf R}$
defines a strongly continuous one-parameter group of unitary
operators on $\mathcal H$ leaving $\Omega_{\beta}$ invariant.

\subsubsection{iii) Construction of an anti-unitary operator $J$}

For
\begin{equation}\label{eqv}
v:=Z[a_1,\tau_1,\ldots,a_n,\tau_n]\in\mathcal V_{\beta},
\end{equation}
$Z\in\mathbf C$, we define
\begin{equation}\label{eqsmalljdef}
jv:=\overline Z[a_n^*,\beta/2-\tau_n,\ldots,a_1^*,\beta/2-\tau_1].
\end{equation}
Equation~(\ref{eqsmalljdef}) is required
 for arbitrary $n$ and hence,
by anti-linearity, defines an anti-linear operator $j$ on all of
$\mathcal V_{\beta}$. Choosing
\[w:=\zeta[b_1,\sigma_1,\ldots,b_m,\sigma_m]\in\mathcal V_{\beta},\ 
\zeta\in\mathbf C,\]
we observe that, by (\ref{eqsmalljdef}) and (\ref{eqproddef}),
\[\langle jv,jw\rangle=\overline Z\zeta\phi_{\beta}(a_n^*,
\beta/2-\tau_n,\ldots,a_1^*,\beta/2-\tau_1,b_1,\beta/2+\sigma_1,
\ldots,b_m,\beta/2+\sigma_m)\]
\[\stackrel{\mbox{(P2)}}{=}\overline Z\zeta\phi_{\beta}(a_n^*,
-\tau_n,\ldots,a_1^*,-\tau_1,b_1,\sigma_1,\ldots,b_m,\sigma_m)\]
\[\stackrel{\mbox{(P3)}}{=}\zeta\overline Z\phi_{\beta}(
b_1,\sigma_1,\ldots,b_m,\sigma_m,a_n^*,\beta-\tau_n,\ldots,
a_1^*,\beta-\tau_1)\]
\begin{equation}\label{eqsmalljantiunit}
\stackrel{\mbox{(\ref{eqproddef})}}{=}
\langle w,v\rangle.
\end{equation}
Thus, if $v\in\mathcal N_{\beta}$,
\[\langle jv,jv\rangle=\langle v,v\rangle=0,\]
i.e.,
\begin{equation}
jv\in\mathcal N_{\beta}.
\end{equation}
This observation enables us to define an anti-linear
operator $J$ on $\mathcal D_{\beta}$ by setting
\begin{equation}\label{eqjconst}
J\Phi(v):=\Phi(jv).
\end{equation}
Then,
\[\langle J\Phi(v),J\Phi(w)\rangle=
\langle\Phi(jv),\Phi(jw)\rangle\]
\[=\langle jv,jw\rangle\]
\[\stackrel{\mbox{(\ref{eqsmalljantiunit})}}{=}
\langle w,v\rangle\]
\begin{equation}
=\langle\Phi(w),\Phi(v)\rangle,
\end{equation}
i.e., $J$ is {\em anti-unitary}.

Next, we note that, for $v$ as in (\ref{eqv}),
\[j(v_{\tau})=\overline Z[a_n^*,\beta/2-\tau_n-\tau,\ldots,
a_1^*,\beta/2-\tau_1-\tau]\]
\begin{equation}
=(jv)_{-\tau},
\end{equation}
for $\tau\in(-\epsilon_-(v),\epsilon_+(v))$. It then follows from
(\ref{eqgammadef}) and (\ref{eqjconst}) that, for $\Psi\in\mathcal D
_{\beta}$ and $\tau\in(-\epsilon_-(\Psi),\epsilon_+(\Psi))$,
\begin{equation}
J\Gamma_{\tau}\Psi=\Gamma_{-\tau}J\Psi.
\end{equation}
Since $J$ is anti-unitary, and by (\ref{eqgammal}),
\begin{equation}\label{eqjcomml}
Je^{itL}=e^{itL}J,\mbox{ or }JL=-LJ.
\end{equation}
Finally,
\begin{equation}
J\Omega_{\beta}=J\Phi([\emptyset])=\Phi([\emptyset])=\Omega_{\beta}.
\end{equation}

\subsection{Proof of Part (2) of the Main Theorem}

To prove part (2) of our Main Theorem, we must assume
that the imaginary-time Green functions (TOGF's), $\phi_{\beta}$,
not only obey Properties (P1)--(P4) of section~\ref{secfprop}, but,
in addition, Property~(P$^*$). In particular, we shall henceforth
assume that $\mathcal S$ is a $C^*$-algebra.

\subsubsection{i) Construction of a $\star$-representation $\lambda$
and an anti-representation $\rho$ of $\mathcal S$ on $\mathcal H_{\beta}$}

Thanks to Property (P$^*$), in particular (P$^*$i), we may define the
linear spaces
\[\tilde{\mathcal V}_{\beta}:=\bigoplus_{n=0}^{\infty}
\tilde{\mathcal V}_{\beta}^{(n)},\]
where
\[\tilde{\mathcal V}_{\beta}^{(n)}:=\left\{\sum_iZ_i
[a_1^i,\tau_1^i,\ldots,a_n^i,\tau_n^i]\right\},\]
with $Z_i\in\mathbf C$, $a_1^i,\ldots,a_n^i\in\mathcal S$, and
\begin{equation}
0\le\tau_1^i\le\tau_2^i\le\cdots\le\tau_n^i\le\beta/2,
\end{equation}
for all $i$; $\tilde{\mathcal V}_{\beta}^{(0)}=
\mathcal V_{\beta}^{(0)}=\mathbf C$. Note that, thanks to Property
(P$^*$ii),
\[[a_1,\tau_1,\ldots,a_j,\tau_j,a_{j+1},\tau_j,\ldots,a_n,\tau_n]\]
\begin{equation}\label{eqident}
\equiv[a_1,\tau_1,\ldots,a_j\cdot a_{j+1},\tau_j,\ldots,a_n,\tau_n]
\end{equation}
must be {\em identified}, for $\tau_{j+1}=\tau_{j}$, for arbitrary
$j$. Obviously, the space $\tilde{\mathcal V}_{\beta}$ contains the
space $\mathcal V_{\beta}$ defined in (\ref{eqvdef}).

For $a\in\mathcal S$ and $v:=[a_1,\tau_1,\ldots,a_n,\tau_n]\in
\tilde{\mathcal V}_{\beta}$, we define
\begin{equation}
av:=[a,0,a_1,\tau_1,\ldots,a_n,\tau_n]\in\tilde{\mathcal V}_{\beta},
\end{equation}
and
\begin{equation}
va^*:=[a_1,\tau_1,\ldots,a_n,\tau_n,a^*,\beta/2]\in\tilde{\mathcal V}_{\beta}.
\end{equation}
These definitions can  be extended to all of $\tilde{\mathcal V}_{\beta}$
by linearity. Let $\tilde{\mathcal N}_{\beta}$ denote the kernel of the
inner product $\langle\cdot,\cdot\rangle$ on $\tilde{\mathcal V}_{\beta}$,
defined as in (\ref{eqproddef}), (\ref{eqndef}) (An example of a vector
in $\tilde{\mathcal N}_{\beta}$ is the difference of the two vectors
in (\ref{eqident})). By (\ref{eqproddef}),
\begin{equation}
\langle va^*,w\rangle=\langle v,wa\rangle,
\end{equation}
and, using the KMS condition (Property (P3)),
\begin{equation}\label{eqastar}
\langle av,w\rangle=\langle v,a^*w\rangle,
\end{equation}
for arbitrary $v$ and $w$ in $\tilde{\mathcal V}_{\beta}$.
These equations and the Cauchy-Schwarz inequality show 
that $\tilde{\mathcal N}_{\beta}$ is a {\em two-sided ideal}
under left- and right multiplication by elements of $\mathcal S$.
This permits us to define a dense, linear subspace, 
$\tilde{\mathcal D}_{\beta}$, of $\mathcal H_{\beta}$ and, for
$a\in\mathcal S$, linear operators $\lambda(a)$ and $\rho(a)$
on $\tilde{\mathcal D}_{\beta}$ by setting
\begin{equation}
\tilde{\mathcal D}_{\beta}:=\Phi(\tilde{\mathcal V}_{\beta})
=\tilde{\mathcal V}_{\beta}\bmod\tilde{\mathcal N}_{\beta},
\end{equation}
and
\begin{equation}\label{eqlambdarhodef}
\lambda(a)\Phi(v):=\Phi(av),\ \rho(a)\Phi(v):=\Phi(va^*),
\end{equation}
for arbitrary $v\in\tilde{\mathcal V}_{\beta}$. We note that
$\lambda(\cdot)$ is {\em linear}, while $\rho(\cdot)$ is
{\em anti-linear} on $\mathcal S$. Property (P$^*$ii) shows that, 
for arbitrary $a$ and $b$ in $\mathcal S$,
\begin{equation}\label{eqlambdarhohomo}
\lambda(a)\cdot\lambda(b)=\lambda(a\cdot b),\
\rho(a)\cdot\rho(b)=\rho(a\cdot b),
\end{equation}
on the domain $\tilde{\mathcal D}_{\beta}$. Further important
properties of $\lambda$ and $\rho$ are described in the following
lemma.

\begin{lemma}\label{lemmalambdaprop}
(1) For arbitrary $a\in\mathcal S$,
\begin{equation}
\rho(a)\Psi=J\lambda(a)J\Psi,\ \Psi\in\tilde{\mathcal D}_{\beta},
\end{equation}
where $J$ is the anti-unitary operator defined in
(\ref{eqsmalljdef}), (\ref{eqjconst});

(2)
\[\langle\lambda(a)\Psi,\tilde{\Psi}\rangle=
\langle\Psi,\lambda(a^*)\tilde{\Psi}\rangle,\]
for arbitrary $\Psi$ and $\tilde{\Psi}$ in $\tilde{\mathcal D}_{\beta}$,
i.e.,
\begin{equation}
\lambda(a)^*\supseteq\lambda(a^*);
\end{equation}

(3) $\lambda(a)$ extends to a bounded operator on $\mathcal H_{\beta}$
with
\begin{equation}
\|\lambda(a)\|\le\|a\|.
\end{equation}
\end{lemma}

\textbf{Remark:} By (1), parts (2) and (3) also hold for
$\rho(a)$, instead of $\lambda(a)$.

\textbf{Proof:} (1) For $v=[a_1,\tau_1,\ldots,a_n,\tau_n]\in
\tilde{\mathcal V}_{\beta}$,
\[J\lambda(a)J\Phi(v)=J\lambda(a)\Phi(jv)\]
\[=J\lambda(a)\Phi[a_n^*,\beta/2-\tau_n,\ldots,a_1^*,\beta/2-\tau_1]\]
\[=J\Phi[a,0,a_n^*,\beta/2-\tau_n,\ldots,a_1^*,\beta/2-\tau_1]\]
\[=\Phi[a_1,\tau_1,\ldots,a_n,\tau_n,a^*,\beta/2]\]
\[=\Phi(va^*)\]
\begin{equation}
=\rho(a)\Phi(v),
\end{equation}
by (\ref{eqjconst}), (\ref{eqsmalljdef}) and (\ref{eqlambdarhodef}).
Part (1) then follows by linearity.

Part (2) is an immediate consequence of (\ref{eqastar}). Here are
some details: Let $v$ be as above and $\tilde v:=[b_1,\sigma_1,
\ldots,b_m,\sigma_m]\in\tilde{\mathcal V}_{\beta}$. We set
$\Psi:=\Phi(v)$, $\tilde{\Psi}:=\Phi(\tilde v)$. Then, using
(\ref{eqlambdarhodef}), (\ref{eqproddef}) and the KMS condition
(P3),
\[\langle\lambda(a)\Psi,\tilde{\Psi}\rangle=\phi_{\beta}
(a,0,a_1,\tau_1,\ldots,a_n,\tau_n,b_m^*,\beta-\sigma_m,\ldots,b_1^*,
\beta-\sigma_1)\]
\[=\phi_{\beta}(a_1,\tau_1,\ldots,a_n,\tau_n,b_m^*,\beta-\sigma_m,\ldots,
b_1^*,\beta-\sigma_1,a,\beta)\]
\[=\langle\Psi,\lambda(a^*)\tilde{\Psi}\rangle.\]

It remains to prove part (3). By the Cauchy-Schwarz inequality, part (2)
and (\ref{eqlambdarhohomo}),
\[|\langle\lambda(a)\Psi,\tilde{\Psi}\rangle|^2\le
\langle\lambda(a)\Psi,\lambda(a)\Psi\rangle
\langle\tilde{\Psi},\tilde{\Psi}\rangle\]
\[=\langle\lambda(a^*a)\Psi,\Psi\rangle
\langle\tilde{\Psi},\tilde{\Psi}\rangle\]
\[\le\langle\lambda(a^*a)\Psi,\lambda(a^*a)\Psi\rangle^{1/2}
\langle\Psi,\Psi\rangle^{1/2}
\langle\tilde{\Psi},\tilde{\Psi}\rangle\]
\[=\langle\lambda((a^*a)^2)\Psi,\Psi\rangle^{1/2}
\langle\Psi,\Psi\rangle^{1/2}
\langle\tilde{\Psi},\tilde{\Psi}\rangle\]
\[\le\langle\lambda((a^*a)^2)\Psi,\lambda((a^*a)^2)\Psi\rangle^{1/4}
\langle\Psi,\Psi\rangle^{3/4}
\langle\tilde{\Psi},\tilde{\Psi}\rangle\]
\[\le\cdots\]
\[\le\langle\lambda((a^*a)^{2^N})\Psi,\Psi\rangle^{2^{-N}}
\langle\Psi,\Psi\rangle^{1-2^{-N}}
\langle\tilde{\Psi},\tilde{\Psi}\rangle,\]
for all $N=1,2,3,\ldots$.

Next, we note that
\[|\langle\lambda((a^*a)^{2^N})\Psi,\Psi\rangle|
=|\phi_{\beta}((a^*a)^{2^N},0,a_1,\tau_1,\ldots,a_n,\tau_n,
a_n^*,\beta-\tau_n,\ldots,a_1^*,\beta-\tau_1)|\]
\[\le\|(a^*a)^{2^N}\|\prod_{j=1}^n(\|a_j\|\cdot\|a_j^*\|)\]
\[\|a\|^{2^{N+1}}\prod_{j=1}^n\|a_j\|^2,\]
by Property (P$^*$iv). We
have used that $\|(\cdot)\|$ is a $C^*$-norm.

By letting $N$ tend to $\infty$, we find that 
\[|\langle\lambda(a)\Psi,\tilde{\Psi}\rangle|\le
\|a\|\langle\Psi,\Psi\rangle^{1/2}\langle\tilde{\Psi},
\tilde{\Psi}\rangle^{1/2},\]
from which part (3) follows by (anti-)linearity in $\Psi$,
$\tilde{\Psi}$, resp.\hfill$\diamondsuit$

\medskip\medskip

We define a linear subspace $\mathcal D_{\beta}^+\subset
\tilde{\mathcal D}_{\beta}$ of $\mathcal H_{\beta}$ by
\begin{equation}
\mathcal D_{\beta}^+:=\{\Phi([a,\beta/2])|a\in\mathcal S\}.
\end{equation}
To each vector $\Psi\in\mathcal H_{\beta}$, we associate an
operator $\hat{\Psi}:\mathcal D_{\beta}^+\rightarrow
\mathcal H_{\beta}$, by setting
\begin{equation}\label{eqpsihatdef}
\hat{\Psi}\Phi([a,\beta/2]):=\rho(a^*)\Psi.
\end{equation}
Clearly,
\begin{equation}\label{eqpsihatomega}
\hat{\Psi}\Omega_{\beta}=\Psi,
\end{equation}
and (\ref{eqpsihatdef}), (\ref{eqpsihatomega}) show
that if
\begin{equation}\label{eqpsihat1ique1}
\tilde{\Psi}=\hat{\tilde{\Psi}}\Omega_{\beta}=
\hat{\Psi}\Omega_{\beta}=\Psi,
\end{equation}
then
\begin{equation}\label{eqpsihat1ique2}
\hat{\tilde{\Psi}}=\hat{\Psi},
\end{equation}
as operators on $\mathcal D_{\beta}^+$.

\begin{lemma}\label{lemmalambdacommrho}
For arbitrary $a,b$ in $\mathcal S$ and real numbers $t,s$,
\begin{equation}\label{eqlambdacommrho}
[e^{itL}\lambda(a)e^{-itL},e^{isL}\rho(b)e^{-isL}]=0,
\end{equation}
where $L$ is the Liouvillian constructed in section~\ref{secproofpart1};
see equation~(\ref{eqgammal}).
\end{lemma}

\textbf{Remark:} Lemmas \ref{lemmalambdaprop} and~\ref{lemmalambdacommrho}
show that $\mathcal H_{\beta}$ is a {\em bi-module} for the
$C^*$-algebra, $\mathcal A$, generated by
\[\left\{e^{itL}\lambda(a)e^{-itL}|a\in\mathcal S,\ t\in\mathbf R\right\}.\]

\textbf{Proof:} Since $\{\exp(itL)\}_{t\in\mathbf R}$ is a 
one-parameter unitary group, it is enough to prove
(\ref{eqlambdacommrho}) for $s=0$.
Let $\tilde{\Psi}\in\mathcal H_{\beta}$. By unitarity of $\exp(itL)$
and part (3) of Lemma~\ref{lemmalambdaprop},
\[\Psi:=e^{itL}\lambda(a)e^{-itL}\tilde{\Psi}\in\mathcal H_{\beta}.\]
Using (\ref{eqpsihat1ique1}) and (\ref{eqpsihat1ique2}),
it is not hard to show that
\[\hat{\Psi}=e^{itL}\lambda(a)e^{-itL}\hat{\tilde{\Psi}}.\]
This equality and (\ref{eqpsihatdef}) then yield
\[\rho(b)\Psi=\hat{\Psi}\Phi([b^*,\beta/2])\]
\[=e^{itL}\lambda(a)e^{-itL}\hat{\tilde{\Psi}}\Phi([b^*,\beta/2])\]
\[=e^{itL}\lambda(a)e^{-itL}\rho(b)\tilde{\Psi},\]
which proves (\ref{eqlambdacommrho}) for $s=0$.\hfill$\diamondsuit$

\medskip\medskip

We conclude this section with a comment on the KMS condition at
real time. For $a$ and $b$ in $\mathcal S$, $t\in\mathbf R$, we
have that
\[\langle e^{itL}\lambda(a)e^{-itL}\lambda(b)
\Omega_{\beta},\Omega_{\beta}\rangle\]
\[=\langle 
\lambda(b)\Omega_{\beta},e^{itL}\lambda(a^*)\Omega_{\beta}\rangle\]
\[=\langle Je^{itL}\lambda(a^*)
\Omega_{\beta},J\lambda(b)\Omega_{\beta}\rangle\]
\[=\langle 
e^{itL}e^{\beta L/2}\lambda(a)\Omega_{\beta},
e^{\beta L/2}\lambda(b^*)\Omega_{\beta}\rangle,\]
by (\ref{eqsmalljdef}), (\ref{eqjconst}) and (\ref{eqgammadef}).
This implies that
\[F_{ab}(t):=\langle e^{itL}\lambda(a)e^{-itL}\lambda(b)
\Omega_{\beta},\Omega_{\beta}\rangle\]
is the boundary value of a function $F_{ab}(z)$ analytic in
$z$ on the strip $\{z|-\beta<\mathrm{Im}z<0\}$, which is the 
KMS condition! In the next subsection, we use somewhat more
sophisticated arguments of this type to reconstruct {\em all}
real-time Green functions from TOGF's, $\phi_{\beta}$,
by analytic continuation in the time variables.

\subsubsection{ii) Back to Real Times}

In this subsection, we show that if a set of TOGF's, 
$\phi_{\beta}(a_1,\tau_1,\ldots,a_n,\tau_n)$, $a_i\in\mathcal S$,
for all $i$, $(\tau_1,\ldots,\tau_n)\in\overline{\mathbf T^n_<}$ 
(see~(\ref{eqtdef})), have Properties (P1)--(P4) and (P$^*$) of
section~\ref{secfprop}, then they are the restrictions of functions
$F_{\beta}(a_1,z_1,\ldots,a_n,z_n)$, analytic in $(z_1,\ldots,z_n)$ 
on the tubular domain $\overline{\mathcal T_n}$ defined in equation 
(\ref{eqfanal}), to the region
\[(z_1,\ldots,z_n)=(i\tau_1,\ldots,i\tau_n),\ 
(\tau_1,\ldots,\tau_n)\in\overline{\mathbf T^n_<}.\]
Real-time Green functions are then obtained as the boundary
values of the functions $F_{\beta}(a_1,z_1,\ldots,a_n,z_n)$
when $z_i$ tends to the real axis, for all $i=1,\ldots,n$.
Our results in this subsection will complete the proof of our
Main Theorem, stated in section~\ref{secmainthm}.

To start with, we note that 
\begin{equation}
\lambda(a_1)\Omega_{\beta}=\lambda(a_1)\Phi([\emptyset])=
\Phi([a_1,0])\in\mathcal H_{\beta},
\end{equation}
for all $a_1\in\mathcal S$. Furthermore, by (\ref{eqgammal})
and (\ref{eqgammadef}),
\[e^{\tau_1L}\lambda(a_1)\Omega_{\beta}=
e^{\tau_1L}\Phi([a_1,0])\]
\begin{equation}
=\Phi([a_1,\tau_1])\in\mathcal H_{\beta},
\end{equation}
for $0\le\tau_1\le\beta/2$. Since $\{\exp(itL)\}_{t\in\mathbf R}$
is a one-parameter {\em unitary} group on $\mathcal H_{\beta}$,
\begin{equation}\label{eq1vec}
e^{z_1L}\lambda(a_1)\Omega_{\beta}=e^{it_1L}\Phi([a_1,\tau_1])
\in\mathcal H_{\beta},
\end{equation}
for $z_1=\tau_1+it_1$, $0\le\tau_1\le\beta/2$, $t_1\in\mathbf R$,
and the left hand side of (\ref{eq1vec}) is {\em holomorphic} in
$z_1$, for $0<\mathrm{Re}z_1\equiv\tau_1<\beta/2$. Furthermore,
\[\left\|e^{z_1L}\lambda(a_1)\Omega_{\beta}\right\|^2=
\langle\Phi([a_1,\tau_1]),\Phi([a_1,\tau_1])\rangle\]
\[=\phi_{\beta}(a_1,\tau_1,a_1^*,\beta-\tau_1)\]
\begin{equation}
\le\|a_1\|^2,
\end{equation}
by Property (P$^*$iv).

Part (3) of Lemma \ref{lemmalambdaprop} then shows that
\begin{equation}\label{eq2vec}
\Psi_{a_2a_1}(z_1):=\lambda(a_2)e^{z_1L}\lambda(a_1)\Omega_{\beta}
\end{equation}
is a holomorphic $\mathcal H_{\beta}$-valued function of $z_1$,
for $0<\mathrm{Re}z_1<\beta/2$, with
\begin{equation}\label{eq2vecbound}
\|\Psi_{a_2a_1}(z_1)\|\le\|a_2\|\,\|a_1\|,
\end{equation}
for $0\le\mathrm{Re}z_1\le\beta/2$.

The idea is now to proceed {\em inductively}, showing that
$\Psi_{a_2a_1}(z_1)$ is in the domain of definition of the
{\em unbounded} operator $\lambda(a_3)\exp(z_2L)$, as long
as $0\le\mathrm{Re}z_2\le\beta/2-\mathrm{Re}z_1$, etc.
The {\em induction hypothesis} is

$\mathbf{[A_{n-1}]}$ {\em For arbitrary $a_1,\ldots,a_n$ in $\mathcal S$,
\begin{equation}\label{eqnvecdef}
\Psi_{a_n\ldots a_1}(z_{n-1},\ldots,z_1):=\lambda(a_n)e^{z_{n-1}L}
\lambda(a_{n-1})\cdots\lambda(a_2)e^{z_1L}\lambda(a_1)\Omega_{\beta}
\end{equation}
is a vector in $\mathcal H_{\beta}$, for all $(z_1,\ldots,z_{n-1})
\in\overline T_{n-1}^{(\beta)}$,
where
\begin{equation}
T_{n-1}^{(\beta)}:=\{(z_1,\ldots,z_{n-1})|\mathrm{Re}z_i>0,\forall i,
\sum_{i=1}^{n-1}\mathrm{Re}z_i<\beta/2\};
\end{equation}
it is holomorphic in $(z_1,\ldots,z_n)\in T_{n-1}^{(\beta)}$
and, on $\overline T_{n-1}^{(\beta)}$, is bounded in norm by
\begin{equation}\label{eqnvecbound}
\|\Psi_{a_n\ldots a_1}(z_{n-1},\ldots,z_1)\|\le
\prod_{j=1}^{n}\|a_j\|.
\end{equation}}

In (\ref{eq2vec}), (\ref{eq2vecbound}), 
$[A_1]$ has been established. We shall now carry out the

\textbf{Induction Step: $\mathbf{[A_{n-1}]\Rightarrow[A_n]}$, 
$\mathbf{\forall n}$.}

Let $\chi_N$ be the characteristic function of the interval $[-N,N]$.
Then,
\[\chi_N(L)e^{zL}=e^{zL}\chi_N(L)\]
is an {\em entire} operator-valued function of $z$,
{\em bounded in norm} by $\exp(N|\mathrm{Re}z|)$. Thus, 
$[A_{n-1}]$ implies that the vectors
\begin{equation}
\Pi^{(N)}_{a_n\ldots a_1}(z_{n-1},\ldots,z_1):=\chi_N(L)
e^{(\beta/2-\sum_{i=1}^{n-1}z_i)L}\Psi_{a_n\ldots a_1}(z_{n-1},\ldots,z_1)
\end{equation}
are well defined, for all $(z_1,\ldots,z_{n-1})\in\overline
T_{n-1}^{(\beta)}$,
and depend holomorphically on $(z_1,\ldots,z_{n-1})$, for
$(z_1,\ldots,z_{n-1})\in T_{n-1}^{(\beta)}$, for all $N<\infty$.
For $z_i=\tau_i$ non-negative, for $i=1,\ldots,n-1$, with
$\sum_{i=1}^{n-1}\tau_i\le\beta/2$,
\[\Pi^{(N)}_{a_n\ldots a_1}(\tau_{n-1},\ldots,\tau_1)=
\chi_N(L)e^{(\beta/2-\sum_{i=1}^{n-1}\tau_i)L}\lambda(a_n)
\left(\prod_{i=n-1}^1e^{\tau_iL}\lambda(a_i)\right)
\Omega_{\beta}\]
\[=\chi_N(L)\Phi\left(\left[a_n,\beta/2-\sum_{i=1}^{n-1}\tau_i,a_{n-1},
\beta/2-\sum_{i=1}^{n-2}\tau_i,\ldots,a_1,\beta/2\right]\right)\]
\[=\chi_N(L)J\Phi\left(\left[
a_1^*,0,a_2^*,\tau_1,\ldots,a_n^*,\sum_{i=1}^{n-1}\tau_i\right]\right)\]
\[=\chi_N(L)J\left(\prod_{i=1}^{n-1}\lambda(a_i^*)e^{\tau_i L}\right)
\lambda(a_n^*)\Omega_{\beta}\]
\begin{equation}\label{eqpicalc}
=\chi_N(L)J\Psi_{a_1^*\ldots a_n^*}(\tau_1,\ldots,\tau_{n-1}),
\end{equation}
by (\ref{eqsmalljdef}), (\ref{eqjconst}) and (\ref{eqnvecdef}).
The induction hypothesis $[A_{n-1}]$ tells us that
$\Psi_{a_1^*\ldots a_n^*}$ $(\overline z_1,\ldots,\overline z_{n-1})$
is holomorphic in  $(\overline z_{n-1},\ldots,\overline z_1)\in
T_{n-1}^{(\beta)}$ and bounded in norm by $\prod_{i=1}^n\|a_i\|$,
for $(\overline z_{n-1},\ldots,\overline z_1)\in\overline T_{n-1}^{(\beta)}$.
Since $J$ is an {\em anti-unitary} operator,
\begin{equation}
J\Psi_{a_1^*\ldots a_n^*}(\overline z_1,\ldots,\overline z_{n-1})
\end{equation}
is {\em holomorphic} in $(z_1,\ldots,z_{n-1})\in T_{n-1}^{(\beta)}$,
and
\begin{equation}\label{eqjpsibound}
\|J\Psi_{a_1^*\ldots a_n^*}(\overline z_1,\ldots,\overline z_{n-1})\|
\le\prod_{i=1}^n\|a_i\|,
\end{equation}
for $(z_1,\ldots,z_{n-1})\in\overline T_{n-1}^{(\beta)}$, by 
(\ref{eqnvecbound}). If $z_i$ is non-negative for $i=1,\ldots,n-1$,
and $\sum_{i=1}^{n-1}z_i\le\beta/2$, then (\ref{eqpicalc})
shows that
\begin{equation}\label{eqpipsi}
\Pi^{(N)}_{a_n\ldots a_1}(z_{n-1},\ldots,z_1)=\chi_N(L)J
\Psi_{a_1^*\ldots a_n^*}(\overline z_1,\ldots,\overline z_{n-1}).
\end{equation}
Since the left hand side and the right hand side of (\ref{eqpipsi})
are {\em holomorphic} $\mathcal H_{\beta}$-valued functions of
$(z_1,\ldots,z_{n-1})\in T_{n-1}^{(\beta)}$, equation (\ref{eqpipsi})
holds for all $(z_1,\ldots,z_{n-1})\in T_{n-1}^{(\beta)}$, for
all $N<\infty$, and, with (\ref{eqjpsibound}) and (\ref{eqpicalc}),
and using that
$\|\chi_N(L)\|=1$, we find that
\begin{equation}\label{eqpibound}
\|\Pi^{(N)}_{a_n\ldots a_1}(z_{n-1},\ldots,z_1)\|\le
\prod_{i=1}^{n}\|a_i\|,
\end{equation}
{\em uniformly} in $N<\infty$. Since $\exp[(\beta/2-\sum_{i=1}^{n-1}z_i)L]$
is a {\em closed} operator, and
\[\mbox{s-}\!\!\!\!\lim_{N\rightarrow\infty}\chi_N(L)
\Psi_{a_n\ldots a_1}(z_{n-1},\ldots,z_1)=
\Psi_{a_n\ldots a_1}(z_{n-1},\ldots,z_1),\]
\[\mbox{s-}\!\!\!\!\lim_{N\rightarrow\infty}\chi_N(L)
J\Psi_{a_1^*\ldots a_n^*}(\overline z_1,\ldots,\overline z_{n-1})=
J\Psi_{a_1^*\ldots a_n^*}(\overline z_1,\ldots,\overline z_{n-1}),\]
for $(z_1,\ldots,z_{n-1})\in\overline T_{n-1}^{(\beta)}$, by $[A_{n-1}]$,
it follows that
\[\mbox{s-}\!\!\!\!\lim_{N\rightarrow\infty}\Pi^{(N)}_{a_n\ldots a_1}
(z_{n-1},\ldots,z_1)=e^{(\beta/2-\sum_{i=1}^{n-1}z_i)L}
\Psi_{a_n\ldots a_1}(z_{n-1},\ldots,z_1)\]
\begin{equation}\label{eqlimpi}
=J\Psi_{a_1^*\ldots a_n^*}(\overline z_1,\ldots,\overline z_{n-1}),
\end{equation}
for $(z_1,\ldots,z_{n-1})\in\overline T_{n-1}^{(\beta)}$, and
the bound (\ref{eqpibound}) remains true in the limit 
$N\rightarrow\infty$.

Next, we define functions $M^{(N)}(\sigma)$ by
\[M^{(N)}(\sigma):=\left\langle\chi_N(L)e^{\sigma(\beta/2-\sum_{i=1}^{n-1}
\mathrm{Re}z_i)L}\Psi_{a_n\ldots a_1}(z_{n-1},\ldots,z_1)\right.,\]
\begin{equation}\label{eqmdef}
\left.\chi_N(L)e^{\sigma(\beta/2-\sum_{i=1}^{n-1}
\mathrm{Re}z_i)L}\Psi_{a_n\ldots a_1}(z_{n-1},\ldots,z_1)\right\rangle.
\end{equation}
Since $\exp(itL)$ is unitary, the right hand side of (\ref{eqmdef})
does not change if\newline
\mbox{$\exp[\sigma(\beta/2-\sum_{i=1}^{n-1}\mathrm{Re}z_i)L]$}
is replaced by $\exp[\sigma(\beta/2-\sum_{i=1}^{n-1}z_i)L]$
in both arguments of the scalar product. Thus, using $[A_{n-1}]$,
(\ref{eqlimpi}), (\ref{eqpibound}) and that $\|\chi_N(L)\|=1$, we find that
\[0\le M^{(N)}(0)\le\prod_{i=1}^n\|a_i\|^2,\]
and
\begin{equation}
0\le M^{(N)}(1)\le\prod_{i=1}^n\|a_i\|^2.
\end{equation}
For $N<\infty$, $M^{(N)}(\sigma)$ is smooth in $\sigma\in\mathbf R$.
Differentiating $M^{(N)}(\sigma)$ twice in $\sigma$ and using that
$L=L^*$, hence $L^2\ge0$, we conclude that $M^{(N)}(\sigma)$ is a
{\em convex} function of $\sigma$. Thus,
\begin{equation}\label{eqmbound}
0\le M^{(N)}(\sigma)\le\max(M^{(N)}(0),M^{(N)}(1))\le\prod
_{i=1}^n\|a_i\|^2,
\end{equation}
for all $\sigma\in[0,1]$, uniformly in $N$. Inequality
(\ref{eqmbound}) and the induction hypothesis $[A_{n-1}]$ show that
\begin{equation}
\chi_N(L)e^{\tau L}\Psi_{a_n\ldots a_1}(z_{n-1},\ldots,z_1)
\end{equation}
is holomorphic in $(z_1,\ldots,z_{n-1})\in T_{n-1}^{(\beta)}$ and
bounded in norm by $\prod_{i=1}^n\|a_i\|$ on $\overline T_{n-1}^{(\beta)}$,
as long as 
\begin{equation}\label{eqtaurange}
0\le\tau\le\beta/2-\sum_{i=1}^{n-1}\mathrm{Re}z_i,
\end{equation}
{\em uniformly} in $N<\infty$. Using the spectral theorem for $L$
and, in particular, that $\exp(\tau L)$ is a closed operator, we
conclude, similarly to (\ref{eqlimpi}), that
\begin{equation}
\mbox{s-}\!\!\lim_{N\rightarrow\infty}\chi_N(L)e^{\tau L}\Psi
_{a_n\ldots a_1}(z_{n-1},\ldots,z_1)=e^{\tau L}\Psi
_{a_n\ldots a_1}(z_{n-1},\ldots,z_1)
\end{equation}
exists and has the same analyticity- and boundedness properties,
provided (\ref{eqtaurange}) holds. Since $\exp(itL)$ is {\em unitary},
for $t\in\mathbf R$, we conclude that, for $z_n=\tau+it$, with
\begin{equation}
0<\mathrm{Re}z_n=\tau<\beta/2-\sum_{i=1}^{n-1}\mathrm{Re}z_i,
\end{equation}
\begin{equation}
e^{z_nL}\Psi_{a_n\ldots a_1}(z_{n-1},\ldots,z_1)
\end{equation}
is an $\mathcal H_{\beta}$-valued function of $(z_1,\ldots,z_n)$.
It is holomorphic in $(z_1,\ldots,z_n)\in T_n^{(\beta)}$ and bounded
in norm by $\prod_{i=1}^n\|a_i\|$, for all $(z_1,\ldots,z_n)\in
\overline T_n^{(\beta)}$. By part (3) of Lemma~\ref{lemmalambdaprop},
\begin{equation}\label{eqn1vec}
\Psi_{a_{n+1}a_n\ldots a_1}(z_n,\ldots,z_1):=\lambda(a_{n+1})
e^{z_nL}\Psi_{a_n\ldots a_1}(z_{n-1},\ldots,z_1)
\end{equation}
is holomorphic in $(z_1,\ldots,z_n)\in T_n^{(\beta)}$ and
\begin{equation}\label{eqn1vecbound}
\|\Psi_{a_{n+1}a_n\ldots a_1}(z_n,\ldots,z_1)\|\le\prod_{i=1}^{n+1}
\|a_i\|,
\end{equation}
for all $(z_1,\ldots,z_n)\in\overline T_n^{(\beta)}$, for arbitrary
$a_{n+1}\in\mathcal S$. Equation (\ref{eqn1vec}) and inequality
(\ref{eqn1vecbound}) establish $[A_n]$, hence the induction step is
complete.

Next, we set
\[b_j:=a_{n-j+1},\ j=1,\ldots,n,\]
\[z'_{j+1}:=z'_j-iz_{n-j},\ j=1,\ldots,n-1.\]
Then, equation (\ref{eqnvecdef}) implies that 
\begin{equation}\label{eqpsiprod}
\langle\Psi_{a_n\ldots a_1}(z_{n-1},\ldots,z_1),\Omega_{\beta}\rangle
=\langle\prod_{j=1}^ne^{iz'_jL}\lambda(b_j)e^{-iz'_jL}
\Omega_{\beta},\Omega_{\beta}\rangle.
\end{equation}
{\em Real-time Green functions}, as in (\ref{eqgreenfunc}),
are obtained from (\ref{eqpsiprod}) by taking the {\em boundary values}
of these functions when $z'_i$ tends to the real axis, for all
$i=1,\ldots,n$. When
\[iz'_j=\tau_j\in\mathbf R,\ j=1,\ldots,n,\]
with $0<\tau_1<\cdots<\tau_n<\beta/2$, then (\ref{eqpsiprod})
is clearly given by
\[\phi_{\beta}(b_1,\tau_1,\ldots,b_n,\tau_n);\]
see (\ref{eqproddef}), (\ref{eqgammadef}), (\ref{eqgammal}), etc.\ In
order to obtain the Green functions on their {\em maximal domain
of analyticity}, $\mathcal T_n$, see (\ref{eqfanal}), one must
consider scalar products
\[\langle\Psi_{a_k\ldots a_1}(z_{k-1},\ldots,z_1),e^{\overline z_kL}
\Psi_{a_{k+1}^*\ldots a_n^*}(\overline z_{k+1},\ldots,\overline z_{n-1})
\rangle\]
and use that
\[e^{zL}\Omega_{\beta}=\Omega_{\beta}.\]
As a consequence of Properties (P3), (P$^*$iv) and analyticity on the 
tubular domain $\mathcal T_n$, they satisfy the KMS condition
(\ref{eqfkms}).

A $C^*$-dynamical system can be constructed by choosing $\mathcal A$
to be the smallest $C^*$-algebra generated, for example, by all
the operators
\[\left.\left\{\int f(t)e^{itL}\lambda(a)e^{-itL}\right|
f\in C_0(\mathbf R),\ a\in\mathcal S\right\},\]
and noticing that
\[\alpha_t(A):=e^{itL}Ae^{-itL},\ A\in\mathcal A,\]
defines a $\star$-automorphism group of $\mathcal A$.

%% file: sec4.tex
In this final section, we first recall the relationship between the
KMS condition for Lorentz boosts in local, relativistic quantum
field theory (QFT) on Minkowski space, at zero temperature, on the
one hand, and the usual connection between spin and statistics (SSC)
and the PCT theorem, on the other hand. Of course our discussion,
which is an adaptation of one in [\ref{FM}], is based on the deep
results in [\ref{J2},\ref{BW}].

We then recall a generalization of part (1) of our Main Theorem and
of the results in section~\ref{secproofpart1} useful for an
``imaginary-time analysis'' of quantum field theory in some 
non-trivial gravitational backgrounds, in particular on 
Schwarzschild space-time [\ref{S}], de Sitter space [\ref{BEM}]
and on anti-de Sitter space ($AdS$), [\ref{LM}].
Our discussion is based on results in [\ref{LM}] and in [\ref{FOS}]
and is meant to merely recall and illustrate the usefulness of the
general results in these papers. It goes beyond these papers only 
in so far as it includes a general form of the KMS condition.

\subsection{SSC and PCT for Local, Relativistic QFT's on Min\-kowski
Space}\label{secsscpct}

We consider a local, relativistic QFT on Minkowski space
$\mathbf M^d$, at zero temperature. We suppose that this theory
satisfies the Wightman axioms; see [\ref{J},\ref{StW}]. Let
$\mathcal H$ denote the Hilbert space of pure state vectors of
the theory, and $\Omega\in\mathcal H$ the {\em vacuum vector}.
We consider a two-dimensional plane, $\pi$, in $\mathbf M^d$
containing a time-like direction. 
We may choose coordinates, $x^0$,
$x^1$, $\vec x$, on $\mathbf M^d$ such that $\pi$ is the 
01-coordinate plane. Let $\mathcal M$ denote the  self-adjoint
operator on $\mathcal H$ representing  the generator of Lorentz
boosts in $\pi$. Let $\Psi(x^0,x^1,\vec x)$ denote a {\em local
field} of the theory. Lorentz covariance implies that
\begin{equation}
e^{i\alpha\mathcal M}\Psi(x^0,x^1,\vec x)e^{-i\alpha\mathcal M}
=\left(S(\alpha)\Psi\right)(x^0_{\alpha},x^1_{\alpha},\vec x),
\end{equation}
where
\[x^0_{\alpha}=\cosh(\alpha) x^0+\sinh(\alpha) x^1,\]
\begin{equation}
x^1_{\alpha}=\sinh(\alpha) x^0+\cosh(\alpha) x^1,
\end{equation}
and $S$ is a finite-dimensional, projective representation of
the Lorentz group, $L_+^{\uparrow}$, of $\mathbf M^d$.

It is well known that, for a QFT satisfying the Wightman axioms,
the passage from real to purely imaginary times ({\em Wick
rotation}) is possible. Let $\Psi^{\sharp}$ denote either
$\Psi$ or $\Psi^*$, and let
\begin{equation}\label{eqschwinger}
S^{(n)}(\sharp_1,t_1,x^1_1,\vec x_1,\ldots,\sharp_n,t_n,x^1_n,\vec x_n)
\end{equation}
denote the imaginary-time Green- or {\em Schwinger functions} of the
fields $\Psi$, $\Psi^*$, where the arguments $(t_j,x^1_j,\vec x_j)$
are points in {\em Euclidean space}, $t_j$ being the imaginary time
of the $j$th point, and $\sharp_j=\emptyset,\star$, if $\Psi$, $\Psi^*$, 
respectively, is inserted in the $j$th argument, for $j=1,\ldots,n$.

We introduce polar coordinates, $(\tau,r)$, in the $(t,x^1)$-coordinate
plane of $\mathbf E^d$, where $\tau$ is the polar angle, and 
$r\ge0$ the radial variable. Let $\mathcal S$ denote the linear space
of {\em column} vectors
\begin{equation}\label{eqa}
a=\left(\begin{array}{c}f_1\\\vdots\\f_k\end{array}\right)
\end{equation}
of Schwartz-space test functions, $f_{\alpha}(r,\vec x)$, on
$\mathbf R^{d-1}$ with support contained in $\{(r,\vec x)|
r\ge0\}$, denote by $\mathcal S^*$ the space of {\em
row} vectors, $(f_1,\ldots,f_k)$, of test functions, and let
$^*$ be the map from $\mathcal S$ to $\mathcal S^*$ given by
\begin{equation}\label{eqastr}
a^*:=(\overline f_1,\ldots,\overline f_k),
\end{equation}
for $a$ as in 
(\ref{eqa}).
In (\ref{eqa}), (\ref{eqastr}), $k$ is the dimension of the 
(projective) representation $S$ of $L_+^{\uparrow}$ under
which $\Psi$ transforms. For $a_1,\ldots,a_n$ in $\mathcal S$,
we define
\[\phi_{2\pi}(a_1^{\sharp_1},\tau_1,\ldots,a_n^{\sharp_n},\tau_n):=\]
\[\int S^{(n)}\left(\sharp_1,t(\tau_1,r_1),x^1(\tau_1,r_1),\vec x_1,
\ldots,\right.\]
\begin{equation}\label{eqdefinephi}
\left.\sharp_n,t(\tau_n,r_n),x^1(\tau_n,r_n),
\vec x_n\right)\prod_{j=1}^na_j^{\sharp_j}(r_j,\vec x_j)
\mathrm dr_j\mathrm d\vec x_j,
\end{equation}
with
\[t(\tau,r):=r\sin\tau,\ x^1(\tau,r):=r\cos\tau.\]
It follows from the results in [\ref{OS}] that the Green functions
$\phi_{2\pi}(a_1^{\sharp_1},\tau_1,\ldots,a_n^{\sharp_n},\tau_n)$
satisfy Properties (P1), continuity, 
and (P4), reflection positivity, of
section~\ref{secfprop}; see equations (\ref{eqbigpidef}), 
(\ref{eqtaurangenoncoin})
with $\beta=2\pi$!
Property (P2) (translation invariance) must be replaced by

(P2') {\em Rotation Invariance}
\begin{equation}
\phi_{2\pi}(a_1^{\sharp_1},\tau_1,\ldots,a_n^{\sharp_n},\tau_n)
=\phi_{2\pi}(a_1^{\sharp_1}(\tau),\tau_1+\tau,\ldots,
a_n^{\sharp_n}(\tau),\tau_n+\tau),
\end{equation}
where
\begin{equation}\label{eqaastar}
a(\tau)=R(\tau)a,\ a^*(\tau)=(R(-\tau)a)^*=(a(-\tau))^*,
\end{equation}
and $R$ is the $k$-dimensional, projective representation of the
group of rotations of $\mathbf E^d$ obtained from the representation
$S$ by analytic continuation in the rapidity, $\alpha$. If $S$ is irreducible,
then
\begin{equation}\label{eqrtwopi}
R(\tau=2\pi)=e^{i2\pi s_{\Psi}},
\end{equation}
where $s_{\Psi}$ is the {\em ``spin''} of the field $\Psi$.
It is well known that
\begin{equation}
s_{\Psi}=0,\frac{1}{2} \bmod\mathbf Z,\mbox{ for $d\ge3$},
\end{equation}
while
\begin{equation}
s_{\Psi}=[0,1) \bmod \mathbf Z,\mbox{ for $d=2$}.
\end{equation}

We are interested in understanding whether a property similar to
Property (P3), section~\ref{secfprop}, i.e., the {\em KMS condition},
holds, too. To this end, we first recall that, in QFT, the Green functions
$\phi_{2\pi}(a_1^{\sharp_1},\tau_1,\ldots,a_n^{\sharp_n},\tau_n)$
are defined for {\em arbitrary}, not necessarily ordered, $n$-tuples
$(\tau_1,\ldots,\tau_n)\in \mathbf T^n$ (with $\beta=2\pi$), 
and
\[\phi_{2\pi}(a_1^{\sharp_1},\tau_1,\ldots,a_n^{\sharp_n},\tau_n)\]
\begin{equation}\label{eqphikmsish}
=e^{i\pi\epsilon_{\Psi}j(n-j)}\phi_{2\pi}(a_{j+1}^{\sharp_{j+1}},\tau_{j+1},
\ldots,a_n^{\sharp_n},\tau_n,a_1^{\sharp_1},\tau_1,\ldots,a_j^{\sharp_j},
\tau_j),
\end{equation}
where $\epsilon_{\Psi}$ is the {\em statistics parameter} of $\Psi$, and
\begin{equation}
\epsilon_{\Psi}=0,1,\mbox{ for $d\ge3$},
\end{equation}
with $\epsilon_{\Psi}=0$ corresponding to {\em Bose-} and 
$\epsilon_{\Psi}=1$ corresponding to {\em Fermi} statistics,
while
\begin{equation}
\epsilon_{\Psi}\in[0,2),\mbox{ for $d=2$},
\end{equation}
({\em fractional-}, or {\em braid statistics} [\ref{Fr}]).

\pagebreak
\begin{thm}\label{thmspinstats}
If Properties (P2') (rotation invariance) and (P4) (reflection
positivity) hold, then
\begin{equation}\label{eqssc}
\epsilon_{\Psi}=2s_{\Psi}\bmod2\mathbf Z,
\end{equation}
\end{thm}
i.e., the usual {\em connection between spin and statistics} (SSC) holds.

\textbf{Proof:} Let $a(\tau)$, $a^*(\tau)$ be as in (\ref{eqaastar}).
Reflection positivity, (P4)\linebreak (for $\beta=2\pi$), says that
\[\phi_{2\pi}(a(\tau),\tau,a^*(-\tau),2\pi-\tau)=
\phi_{2\pi}(a(\tau),\tau,(a(\tau))^*,2\pi-\tau)\ge0.\]
Hence,
\[e^{-i2\pi s_{\Psi}}\phi_{2\pi}(a(\tau),\tau,a^*(2\pi-\tau),2\pi-\tau)\]
\[=\phi_{2\pi}(a(\tau),\tau,R(2\pi)^{-1}a^*(2\pi-\tau),2\pi-\tau)\]
\[=\phi_{2\pi}(a(\tau),\tau,a^*(-\tau),2\pi-\tau)\]
\begin{equation}\label{eqsphipos}
\ge0.
\end{equation}
Rotation invariance, (P2'), implies that
\begin{equation}\label{eqphipirot}
\phi_{2\pi}(a(\tau),\tau,a^*(2\pi-\tau),2\pi-\tau)=
\phi_{2\pi}(a(\tau-\pi),\tau-\pi,a^*(\pi-\tau),\pi-\tau).
\end{equation}
By (\ref{eqphikmsish}), we have that
\[\phi_{2\pi}(a(\tau-\pi),\tau-\pi,a^*(\pi-\tau),\pi-\tau)\]
\[=e^{i\pi\epsilon_{\Psi}}\phi_{2\pi}(a^*(\pi-\tau),\pi-\tau,
a(\tau-\pi),\tau-\pi)\]
\[=e^{i\pi\epsilon_{\Psi}}\phi_{2\pi}(a^*(\pi-\tau),\pi-\tau,
R(-2\pi)a(\pi+\tau),\pi+\tau)\]
\begin{equation}\label{eqsepsphi}
=e^{i\pi\epsilon_{\Psi}}e^{-i2\pi s_{\Psi}}
\phi_{2\pi}(a^*(\pi-\tau),\pi-\tau,a(\pi+\tau),\pi+\tau).
\end{equation}
By (P4) and (\ref{eqsphipos}),the product of the second and the
third factors on the right hand side is {\em positive}. Thus,
comparing (\ref{eqsepsphi}) with (\ref{eqphipirot}) and (\ref{eqsphipos}),
we readily find that
\begin{equation}\label{eqseps}
e^{i\pi\epsilon_{\Psi}}=e^{i2\pi s_{\Psi}},
\end{equation}
or
\[\epsilon_{\Psi}=2s_{\Psi}\bmod2\mathbf Z,\]
which completes our proof.\hfill$\diamondsuit$

\medskip\medskip

The heuristic idea behind our proof is captured in the following
formal calculation: For $\tau\in(0,\pi)$,
\[0\le\langle e^{\tau\mathcal M}\Psi(0,a)\Omega,
e^{\tau\mathcal M}\Psi(0,a)\Omega\rangle\]
\[=\langle\Psi(0,a)\Omega,
e^{2\tau\mathcal M}\Psi(0,a)\Omega\rangle,\ 
(\mathcal M=\mathcal M^*)\]
\[=\phi_{2\pi}(a,0,a^*(-2\tau),2\tau)\]
\[=e^{i\pi\epsilon_{\Psi}}\phi_{2\pi}(a^*(-2\tau),2\tau,a,0)\]
\[=e^{i\pi\epsilon_{\Psi}}\langle
\Psi^*\left(2\tau,(R(2\tau)a)^*\right)\Omega,
\Psi^*(0,a^*)\Omega\rangle\]
\[\stackrel{\tau\rightarrow\pi}{=}e^{i\pi\epsilon_{\Psi}}
e^{-i2\pi s_{\Psi}}\langle\Psi^*(0,a^*)\Omega,
\Psi^*(0,a^*)\Omega\rangle,\]
and the last factor on the right hand side is positive, which
yields (\ref{eqseps}).

\textbf{Remark:} The {\em general} SSC for relativistic QFT's
with arbitrarily many local Bose- and Fermi fields in
dimension $d\ge3$ has been established by Araki in [\ref{A3}].
SSC for two-dimensional theories or three-dimensional gauge
theories with {\em braid statistics} has been established in
[\ref{FG}].

Next, we apply Theorem~\ref{thmspinstats} to establish the
$2\pi$-{\em KMS condition} for the {\em Lorentz boosts},
\[\Psi^{\sharp}\mapsto 
e^{i\alpha\mathcal M}\Psi^{\sharp}e^{-i\alpha\mathcal M},\]
at {\em imaginary rapidities} (``times'').

\begin{corl}[KMS]\label{corlkms}
In $d\ge3$ dimensions, the Green functions 
$\phi_{2\pi}(a_1^{\sharp_1},\tau_1,$
$\ldots,a_n^{\sharp_n},\tau_n)$
satisfy the {\em KMS condition}
\[\phi_{2\pi}(a_1^{\sharp_1},\tau_1,\ldots,a_n^{\sharp_n},\tau_n)=\]
\begin{equation}\label{eqphikms}
\phi_{2\pi}(a_{j+1}^{\sharp_{j+1}},\tau_{j+1},\ldots,a_n^{\sharp_n},\tau_n,
a_1^{\sharp_1}(2\pi),\tau_1+2\pi,\ldots,a_j^{\sharp_j}(2\pi),\tau_j+2\pi).
\end{equation}
\end{corl}

\textbf{Proof:} Equation (\ref{eqphikmsish}) tells us that
\[\phi_{2\pi}(a_1^{\sharp_1},\tau_1,\ldots,a_n^{\sharp_n},\tau_n)\]
\[=e^{i\pi\epsilon_{\Psi}j(n-j)}\phi_{2\pi}(a_{j+1}^{\sharp_{j+1}},
\tau_{j+1},\ldots,a_n^{\sharp_n},\tau_n,a_1^{\sharp_1},\tau_1,\ldots,
a_j^{\sharp_j},\tau_j)\]
\begin{equation}\label{eqphikmsproof}
=e^{i\pi\epsilon_{\Psi}j(n-j)}\phi_{2\pi}(a_{j+1}^{\sharp_{j+1}},
\tau_{j+1},\ldots,a_n^{\sharp_n},\tau_n,a_1^{\sharp_1},\tau_1+2\pi,
\ldots,a_j^{\sharp_j},\tau_j+2\pi).
\end{equation}
If $s_{\Psi}\not=0$ ($\epsilon_{\Psi}\not=0$), then $n$ must necessarily
be {\em even} for the Green functions in (\ref{eqphikms}),
(\ref{eqphikmsproof}) to be different from zero. Then,
\[j(n-j)\cong j^2\cong j\pmod{\mathbf Z}.\]
Hence, by theorem~\ref{thmspinstats},
\[e^{i\pi\epsilon_{\Psi}j(n-j)}=e^{i2\pi s_{\Psi}j}.\]
The proof is completed by using (\ref{eqrtwopi}).\hfill$\diamondsuit$

\medskip\medskip

Thanks to Corollary~\ref{corlkms}, we may now define an anti-linear
involution, $J$, as follows: On a vector
\begin{equation}\label{eqpsivec}
\Psi=z\Phi\left([a_1^{\sharp_1}(\tau_1),\tau_1,\ldots,
a_n^{\sharp_n}(\tau_n),\tau_n]\right)
\end{equation}
in the Hilbert space reconstructed from the Schwinger functions
(\ref{eqschwinger}) of a QFT, as in [\ref{OS}], we set
\begin{equation}\label{eqdefinej}
J\Psi:=\overline z\Phi\left([(a_n^{\sharp_n}(\tau_n-\pi))^*,\pi-\tau_n,
\ldots,(a_1^{\sharp_1}(\tau_1-\pi))^*,\pi-\tau_1]\right).
\end{equation}
A variant of the Reeh-Schlieder theorem [\ref{J},\ref{StW}] shows
that vectors of the form (\ref{eqpsivec}) span a dense set in the
Hilbert space of the theory, and a calculation essentially identical
to (\ref{eqsmalljantiunit}), based on using (\ref{eqaastar})
and the KMS condition (\ref{eqphikms}),
proves that $J$ is an {\em anti-unitary involution} (see also
(\ref{eqsmallanti}), (\ref{eqbigjinvol}), below). By (\ref{eqjcomml}),
\begin{equation}\label{eqjcommm}
Je^{i\tau\mathcal M}=e^{i\tau\mathcal M}J,\ \tau\in\mathbf R.
\end{equation}
Equations (\ref{eqdefinej}), (\ref{eqjcommm}) and (\ref{eqdefinephi})
show that $J$ has the interpretation
\begin{equation}
J=\mathrm{P_1CT},
\end{equation}
where P$_1$  represents the {\em spatial reflection}
\begin{equation}\label{eqponedef}
(x^0,x^1,\vec x)\mapsto(x^0,-x^1,\vec x),
\end{equation}
C is {\em charge conjugation}, i.e.,  $\Psi\mapsto\Psi^*$,
and T represents {\em time reversal},
\begin{equation}
(x^0,x^1,\vec x)\mapsto(-x^0,x^1,\vec x).
\end{equation}
If the dimension $d$ is {\em even}, the product of the
reflection (\ref{eqponedef}) with {\em space reflection},
\[\mathrm P:(x^0,x^1,\vec x)\mapsto(x^0,-x^1,-\vec x),\]
has determinant $+1$, hence belongs to $L_+^{\uparrow}$.
Thus, PP$_1$ is always a symmetry of the theory, and hence
\begin{equation}
\Theta:=\mathrm{PCT=PP_1}J
\end{equation}
is always an {\em anti-unitary symmetry} of the theory. This
is Jost's PCT theorem [\ref{J2}].

\textbf{Remarks}

(1) The KMS condition (\ref{eqphikms}) and equation (\ref{eqphikmsish})
(for $n=2$) also imply the SSC, $\epsilon_{\Psi}=2s_{\Psi}\bmod2\mathbf Z$,
{\em without} assuming reflection positivity:
By (\ref{eqphikms}), (\ref{eqaastar}) and (\ref{eqrtwopi}),
\[\phi_{2\pi}(a_1^{\sharp},\tau_1,a_2^{\sharp},\tau_2)=
\phi_{2\pi}(a_2^{\sharp},\tau_2,a_1^{\sharp}(2\pi),\tau_1+2\pi)\]
\begin{equation}\label{eqphispsi}
=e^{i2\pi s_{\Psi}}\phi_{2\pi}(a_2^{\sharp},\tau_2,a_1^{\sharp},\tau_1),
\end{equation}
which when compared with (\ref{eqphikmsish})
proves the SSC, equation (\ref{eqssc}).

(2) Quite clearly, QFT's with {\em braid statistics} in two
or three space-time dimensions require a somewhat more elaborate
analysis, which we will not present here; but see [\ref{FG}].
Suffice it to say that Theorem~\ref{thmspinstats}, suitably
interpreted, remains valid. Our analysis shows that, in three
dimensions, braid statistics and fractional spin do {\em not} 
arise in theories with only point-like localized fields.

\subsection{QFT in some Non-Trivial Gravitational\\ Backgrounds}

Let $X^d$ be a $d$-(complex-)dimensional complex manifold 
equipped with a (symmetric) quadratic form, $\eta$, on the
tangent bundle $\mathrm TX^d$. We assume that $X^d$ contains
two $d$-{\em real}-dimensional submanifolds, $N_L^d$ and
$N_E^d$, such that 
\begin{equation}
\eta_L:=\eta|_{\mathrm TN_L^d}\mbox{ is a Lorentz metric on }N_L^d,
\end{equation}
and
\begin{equation}
\eta_E:=\eta|_{\mathrm TN_E^d}\mbox{ is a Riemannian metric on }N_E^d.
\end{equation}
We shall interpret $N_L^d$ as the {\em space-time} of a physical
system and will be interested in studying local QFT's on $N_L^d$.
Our strategy will be to attempt to construct {\em ``imaginary-time''
Green functions} over the {\em Riemannian slice}, $N_E^d$, of $X^d$
and reconstruct from them data of a local quantum theory on $N_L^d$.
This can be viewed as a general version of the {\em ``Wick rotation''}.
Our analysis is based on results in [\ref{LM},\ref{FOS}]. It actually
does {\em not} make use of the complex manifold $X^d$ --- requiring
$N_E^d$ to have appropriate symmetry properties will suffice! Our
techniques are group-theoretical. The r\^ole of the KMS condition
will be elucidated.

Here are examples of space-times which fit into the context described
above.

(i) \textbf{Complexified Minkowski space:}
\[X^d=\mathbf C^d\]
Points in $X^d$ are denoted by $z=(z^0,\vec z)$; and
\begin{equation}
\eta(z)=-(\mathrm dz^0)^2+(\mathrm d\vec z)^2.
\end{equation}
Then
\[N_L^d=\mathbf M^d=\{z=(x^0,\vec x)|x^0\in\mathbf R,\vec x\in
\mathbf R^{d-1}\},\]
\begin{equation}
N_E^d=\mathbf E^d=\{z=(x^0=it,\vec x)|t\in\mathbf R,\vec x\in
\mathbf R^{d-1}\}.
\end{equation}

(ii) \textbf{de Sitter and AdS:}
We choose
\begin{equation}
X^d:=\{z=(z^0,z^1,\ldots,z^d)\in\mathbf C^{d+1}|
\sum_{j=0}^d(z^j)^2=R^2\},
\end{equation}
for some $R>0$, and $\eta$ to be the restriction of
$\sum_{j=0}^d(\mathrm dz^j)^2$ to $X^d$. Then
\[N_L^d:=dS_R^d=\{z=(ix^0,x^1,\ldots,x^d)|x^j\in\mathbf R,\ j=0,\ldots,d,\]
\[\left.-(x^0)^2+\sum_{j=1}^d(x^j)^2=R^2\right\},\]
\[N_E^d:=S_R^d=\{z=(x^0,x^1,\ldots,x^d)|x^j\in\mathbf R,\ j=0,\ldots,d,\]
\begin{equation}
\left.\sum_{j=0}^d(x^j)^2=R^2\right\}
\end{equation}
are $d$-dimensional {\em de Sitter space} and the $d$-{\em sphere},
respectively. Furthermore,
\[\tilde N_L^d:=AdS^d=\{z=(x^0,ix^1,\ldots,ix^{d-1},x^d)|
x^j\in\mathbf R,\ j=0,\ldots,d,\]
\[\left.(x^0)^2+(x^d)^2-\sum_{j=1}^{d-1}(x^j)^2
=R^2\right\},\]
\[\tilde N_E^d:=\mathbf H^d=\{z=(ix^0,ix^1,\ldots,ix^{d-1},x^d)|
x^j\in\mathbf R,\ j=0,\ldots,d,\]
\begin{equation}\label{eqhyperbolic}
\left.(x^d)^2-\sum_{j=0}^{d-1}(x^j)^2
=R^2\right\}
\end{equation}
are $d$-dimensional {\em anti-de Sitter space} and {\em hyperbolic
space}, respectively.

Obviously, $AdS^d$ is not simply connected. It admits time-like closed
curves. In general, there will be a connection between imaginary-time
Green functions on $\mathbf H^d$ and a quantum theory on the 
(universal) covering space, $\widetilde {AdS^d}$, [\ref{BM}].

Recently, the so-called $AdS$-CFT correspondence has been discovered
[\ref{Mal}] and widely studied. The simplest example of this 
correspondence is one between QFT's on $AdS^2$ and chiral conformal
field theories on a light ray. Here, we just wish to note that
results concerning the passage from imaginary-time Green functions
$\mathbf H^d$ to quantum theory on $\widetilde{AdS^d}$ can be
translated into statements concerning conformal field theory;
see [\ref{LM},\ref{BM}].

Next, we consider examples with the following product structure:
\[X^d=U^k\times Y^{d-k},\ k=1,2,\ldots,\]
where $U^k$ is a subset of a $k$-dimensional complex manifold,
while $Y^{d-k}$ is a $(d-k)$-dimensional, {\em real} manifold,
and
\begin{equation}
N_L^d=U_L^k\times Y^{d-k},\ N_E^d=U_E^k\times Y^{d-k}.
\end{equation}
Here is a concrete example.

(iii) \textbf{Schwarzschild black hole:}
\[X^d=U^2\times S^{d-2},\]
\begin{equation}\label{eqmetricschwarz}
\eta(z)=h(\xi^2)(\mathrm d\xi^2+\xi^2\mathrm d\tau^2)
+k(\xi^2)\mathrm ds^2,
\end{equation}
where $(\xi,\tau)$ are suitable coordinates on $U^2\subset\mathbf C^2$,
and $h$ and $k$ are analytic 
functions of $\xi^2$, positive on the real axis. 
Then
\[N_L^d=U_L^2\times S^{d-2},\]
with
\begin{equation}\label{eqschwarzul}
U_L^2=\{(\xi,\tau=it)|\xi,t\in\mathbf R\}.
\end{equation}
Note that, for $d=4$, the space-time {\em outside the horizon} of a
Schwarzschild black hole (together with its isometric twin) is of the
form (\ref{eqmetricschwarz}), (\ref{eqschwarzul}).
\[N_E^d=U_E^2\times S^{d-2},\]
with
\begin{equation}
U_E^2=\{(\xi,\tau)|\xi\ge0,\ \tau\in\mathbf R/2\pi\mathbf Z\}.
\end{equation}
See e.g.\ [\ref{St}] for background material.

From now on, {\em only} the {\em Riemannian manifold} $N_E^d$ will
be featured, as promised. We must specify the properties of $N_E^d$
needed in our analysis and then check that they are valid in the
examples just considered. We simplify our notation by setting
$N:=N_E^d$, $\eta:=\eta_E=\eta|_{\mathrm TN_E^d}$.

\textbf{Properties of $(N,\eta)$}

(I) \textbf{Reflection symmetry:} $N$ admits an {\em isometric
involution (reflection)}, $r$.  The fixed-point set of $r$
is a submanifol, $M$, of $N$, of co-dimension 1; $M$ is called
the {\em ``equator''} of $N$. It is equipped with the induced
metric.

(II) \textbf{Killing symmetries:} There is a {\em real
symmetric space} $(G,K,\sigma)$, where $G$ is a real,
simply connected Lie group, $\sigma$ is an involutive
homomorphism of $G$, and $K\subset G$ is the fixed-point set
of $\sigma$, with the properties that there is an {\em action,
$\pi$, of $G$ on $N$ generated by Killing vector fields of the
metric $\eta$}, that the equator $M$ is {\em invariant} under
the action of $K$, and
\begin{equation}
r\pi(g)r=\pi(\sigma(g)),\mbox{ for all }g\in G.
\end{equation}
Of course, $K$ is a subgroup of $G$.

It is an easy exercise to check that in all our examples, (i)--(iii),
$N:=N_E^d(\tilde N_E^d)$, $\eta:=\eta_E$ have Properties (I) and (II).
The simplest examples have the following structure:
\[G=U(1)\times K,\mbox{ or }G=\tilde U(1)\times K=\mathbf R\times K;\]
for $g=(e^{i\alpha},k)\in U(1)\times K$,
\[\sigma(e^{i\alpha},k)=(e^{-i\alpha},k).\]
In these examples,
\[X^d=\mathbf C\times M,\ N=N_E^d=\mathbf R\times M,\mbox{ or }
N=S^1\times M,\ N_L^d=i\mathbf R\times M.\]
Examples, where $G=L\times K$, with $L$ some non-abelian Lie group, are
incompatible with the condition that, at zero temperature, the energy
spectrum be contained in $\mathbf R^+$; see [\ref{FOS}].

Following [\ref{FOS}], we next describe the general mathematical 
structure underlying a formulation of local, relativistic
quantum theory at imaginary time. It consists of the following
objects.

(a) A {\em Riemannian manifold $(N,\eta)$} with Properties (I) and
(II), above. We let $(G,K,\sigma)$ denote the symmetric space
appearing in Property (II). (For simplicity, we may assume that
$K$ is a compact subgroup of $G$. This would exclude examples
(i) and $AdS^d\leftrightarrow\mathbf H^d$, equation (\ref{eqhyperbolic}),
above. But these examples {\em are} covered by the results
of [\ref{FOS},\ref{LM}].)

(b) A {\em separable topological vector space}, $\mathcal V$, containing
two isomorphic subspaces, $\mathcal V_+$ and $\mathcal V_-$,
(usually with $\mathcal V_+\cap\mathcal V_-=\{0\}$).

(c) A {\em continuous representation, $\rho$, of the Lie group $G$ on
$\mathcal V$} with the property that, {\em for every} $v\in\mathcal V_{\pm}$,
there exists an open neighbourhood, $U_v$, of the identity element
$e\in G$ such that 
\begin{equation}
\rho(g)v\in\mathcal V_{\pm},\mbox{ for all }g\in U_v,
\end{equation}
and
\begin{equation}
\rho(k)v\in\mathcal V_{\pm},\mbox{ for all }k\in K.
\end{equation}

\pagebreak
(d) An {\em anti-linear involution, $\theta_r$, on $\mathcal V$}
representing the reflection $r$ in Property (I) such that
\begin{equation}
\theta_r\mathcal V_{\pm}=\mathcal V_{\mp},
\end{equation}
and
\begin{equation}\label{eqthetarho}
\theta_r\rho(g)\theta_r=\rho(\sigma(g)),
\end{equation}
for  all $g\in G$.

(e) A {\em bilinear functional, $\phi$, on $\mathcal V_+\times
\mathcal V_-$} with the following properties.

(\~P1) \textbf{Continuity:} $\phi$ is continuous on 
$\mathcal V_+\times\mathcal V_-$ in the product topology of
$\mathcal V\times\mathcal V$.

(\~P2) \textbf{Invariance:} 
\begin{equation}
\phi(\rho(g)v,\rho(g)w)=\phi(v,w),
\end{equation}
for all $g\in U_v\cap U_w$.

(\~P3) \textbf{KMS condition:} Let $h$ be any
element of $G$ such that
\[\sigma(h)=h^{-1},\ \rho(h)\mathcal V_{\pm}\subseteq\mathcal V_{\mp}.\]
Then, for arbitrary $v\in\mathcal V_+$, $w\in\mathcal V_-$,
\begin{equation}\label{eqptildethree}
\phi(v,w)=\phi(\rho(h^{-1})w,\rho(h)v).
\end{equation}

(\~P4) \textbf{Reflection positivity:} For arbitrary 
$v\in\mathcal V_+$,
\begin{equation}
\phi(v,\theta_rv)\ge0.
\end{equation}

The point is that the structure described here enables us to formulate
and prove a {\em generalization} of part (1) of the {\em Main  Theorem}
proven in section~\ref{secrecn}. Our result involves the Lie group, 
$G^*$, {\em dual} to the Lie group $G$ of Killing symmetries of the
manifold $(N,\eta)$. The  group $G^*$ is defined as follows. Let
$\mathbf g$ denote the Lie algebra of $G$ and $\mathbf k$ the Lie algebra
of $K$, the symmetry group of the ``equator'' of $N$. Clearly,
\begin{equation}
[\mathbf k,\mathbf k]\subseteq\mathbf k,
\end{equation}
where $[\cdot,\cdot]$ is the Lie bracket on $\mathbf g$, and
$\mathbf g$ has a decomposition into linear subspaces,
\begin{equation}
\mathbf g=\mathbf k\oplus\mathbf m,
\end{equation}
with the property that
\[[\mathbf k,\mathbf m]\subseteq\mathbf m,\ 
[\mathbf m,\mathbf m]\subseteq\mathbf k,\]
\begin{equation}\label{eqdecompprop}
\sigma|_{\mathbf k}=\mathrm{id},\ \sigma|_{\mathbf m}=-\mathrm{id}.
\end{equation}
The {\em dual} symmetric Lie algebra, $\mathbf g^*$, is defined by
\begin{equation}
\mathbf g^*:=\mathbf k\oplus i\mathbf m.
\end{equation}
By (\ref{eqdecompprop}), $\mathbf g^*$ is again a real Lie algebra.
Let $G^*$ be the {\em simply connected, real} Lie group with Lie
algebra $\mathbf g^*$. We say that $G^*$ is {\em dual} to $G$, and that
$(G^*,K,\sigma)$ is the symmetric space dual to $(G,K,\sigma)$.

The idea is that $G^*$ is the group of Killing symmetries of
{\em ``physical space-time''}, $(N_L,\eta_L)$, associated with
$(N=N_E,\eta=\eta_E)$. One may expect that, usually, $(N_L,\eta_L)$
can  be reconstructed unambiguously from $(N,\eta)$ if $(N,\eta)$
has Properties (I) and (II), with $\mathrm{dim}(\mathbf m)\ge1$
(with $N_L$ assumed to be simply connected).

Let $h$ be an element of $G$ as described in Property (\~P3).
It is easy to see that
\begin{equation}\label{eqhm}
h=\exp M,\mbox{ for some }M\in\mathbf m,
\end{equation}
and, assuming that Property (\~P2) holds, too, that
\begin{equation}\label{eqkhk}
k^{-1}hk\mbox{ satisfies equation (\ref{eqptildethree}),
for all }k\in K.
\end{equation}
Let $\mathbf g_h$ be the Lie subalgebra of $\mathbf g$ on which
the adjoint action of $h$ is trivial, let $\mathbf g_h^*$ be
the corresponding Lie subalgebra of $\mathbf g^*$; and let
$G_h$, $G_h^*$ be the subgroups of $G$ and $G^*$ generated
by $\mathbf g_h$ and $\mathbf g_h^*$, respectively. Clearly
$G_h$ is the subgroup of $G$ commuting with $h$. We note that
$G_h$ contains the one-parameter subgroup 
$\{\exp\tau M|\tau\in\mathbf R\}$.

Henceforth, any element $h$ of $G$ 
as in Property (\~P3)
 is called a {\em KMS-element} of $G$.

We are now prepared to state the main result of this section.
\pagebreak

\begin{thm}\label{thmqft}
Let $(N,\eta)$ have Properties (I) and (II). Furthermore, let
$(N,\eta)$, the associated symmetric space $(G,K,\sigma)$,
$\mathcal V$, $\rho$, $\theta_r$ and $\phi$ be as  described
in points (a) through (e), above, with Properties (\~P1) 
through (\~P4).

These data uniquely determine a separable Hilbert space, $\mathcal H$,
a continuous, unitary representation, $\pi$, of the group $G^*$ on
$\mathcal H$, and, for any KMS-element $h$ of $G$, an anti-unitary
involution, $J_h$, such that
\begin{equation}\label{eqpijcomm}
\pi(g)J_h=J_h\pi(g),\mbox{ for all }g\in G_h^*,
\end{equation}
and
\begin{equation}
\pi(k)J_h=J_{khk^{-1}}\pi(k),
\end{equation}
for all $k\in K$.

\end{thm}

\textbf{Remarks:} (1) With the exception of the statements concerning
the anti-unitary operators, $J_h$, associated with KMS-elements 
$h\in G$, this theorem has been proven, under different hypotheses
on $(G,K,\sigma)$, in [\ref{FOS}] and [\ref{LM}].

(2) The proof of the theorem follows steps i), ii) and iii) of the
proof of part (1) of the Main Theorem of section~\ref{secrecn};
see section~\ref{secproofpart1}.

\subsubsection{i) Construction of Hilbert Space}
An inner product, $\langle\cdot,\cdot\rangle$, on the subspace
$\mathcal V_+\subset\mathcal V$ is defined by
\begin{equation}
\langle v,w\rangle:=\phi(v,\theta_rw);
\end{equation}
see (\~P4). Let $\mathcal N$ denote the kernel of $\langle\cdot,
\cdot\rangle$ in $\mathcal V_+$, and
\begin{equation}
\Phi(v):=v\bmod\mathcal N,\ v\in\mathcal V_+.
\end{equation}
One defines $\mathcal H$ to be the closure of
$\Phi(\mathcal V)\equiv\mathcal V/\mathcal N$ in the norm
determined by $\langle\cdot,\cdot\rangle$, and
\[\langle\Phi(v),\Phi(w)\rangle:=\langle v,w\rangle\]
defines the scalar product on $\mathcal H$.

\subsubsection{ii) Construction of a Unitary Representation,
$\pi$, of $G^*$ on $\mathcal H$}

The representation $\pi$ is defined as follows: For $k\in K$,
\begin{equation}\label{eqpik}
\pi(k)\Phi(v):=\Phi(\rho(k)v).
\end{equation}
It follows directly from Property (\~P2) and (\ref{eqthetarho})
that $\pi(k)$ is a unitary operator. Furthermore, with every 
$M\in\mathbf m$ we associate an operator $\mathcal M$ on 
$\mathcal H$ by setting
\[e^{t\mathcal M}\Phi(v):=\Phi(\rho(e^{tM})v),\]
for $t$ so small that $\exp(tM)\in U_v$.
Note that by (\ref{eqthetarho}), (\ref{eqdecompprop}), Property (\~P2)
and results in [\ref{FOS}], $\mathcal M$ is self-adjoint. Thus,
\[\pi(e^{itM}):=e^{it\mathcal M}\]
defines a one-parameter unitary group. As shown in  [\ref{FOS}] and
[\ref{LM}], under somewhat different hypotheses, $\pi$ defines
a unitary representation of $G^*$ on $\mathcal H$.


\subsubsection{iii) Construction of the anti-unitary involution $J_h$,
$h$ a KMS-element of $G$}

If $h$ is a KMS-element of $G$ and $v\in\mathcal V_+$, we set
\begin{equation}\label{eqsmalljh}
j_hv:=\theta_r\rho(h)v=\rho(h^{-1})\theta_rv.
\end{equation}
Then
\[\langle j_hv,j_hw\rangle=\phi(\theta_r\rho(h)v,\rho(h)w)\]
\[=\phi(\rho(h^{-1})\theta_rv, \rho(h)w)\]
\[=\phi(w,\theta_rv)\]
\begin{equation}\label{eqsmallanti}
=\langle w,v\rangle,
\end{equation}
by (\ref{eqthetarho}) and
Property (\~P3), (\ref{eqptildethree}).
It follows that $\mathcal N$ is invariant under $j_h$, and we
may thus set
\begin{equation}\label{eqbigjphi}
J_h\Phi(v):=\Phi(j_hv),\ v\in\mathcal V_+.
\end{equation}
Equation (\ref{eqsmallanti}) then implies that $J_h$ is
{\em anti-unitary}. Furthermore, by (\ref{eqthetarho}) and
Property (\~P3),\pagebreak
\[J_h^2\Phi(v)=\Phi(j_h^2v)=\Phi(\theta_r\rho(h)\theta_r\rho(h)v)\]
\[=\Phi(\rho(h^{-1})\rho(h)v)\]
\[=\Phi(v),\mbox{ for arbitrary }v\in\mathcal V_+.\]
Hence,
\begin{equation}\label{eqbigjinvol}
J_h^2=\mathbf1.
\end{equation}

Using
(\ref{eqsmalljh}), (\ref{eqbigjphi}), (\ref{eqkhk}) and
(\ref{eqpik}),
we find that
\[\pi(k)J_h\Phi(v)=\Phi(\rho(k)j_hv)\]
\[=\Phi(j_{khk^{-1}}\rho(k)v)\]
\begin{equation}
=J_{khk^{-1}}\pi(k)\Phi(v),
\end{equation}
for all $k\in K$ and all $v\in\mathcal V_+$.

Equation (\ref{eqpijcomm}) easily follows from  the definition
of $G_h$ and $G_h^*$ by using theorems 1 and 3 of [\ref{FOS}].

(3) To come up with an analogue of part (2) of the Main Theorem
stated in section~\ref{secmainthm} and of the results in 
section~\ref{secsscpct} would require introducing more structure.
As an example, let us imagine that $\mathcal V_+$ contains a
linear subspace, $\mathcal V_0$, with
\begin{equation}\label{eqthetavnaught}
\theta_r\mathcal V_0=\mathcal V_0
\end{equation}
(hence $\mathcal V_0\subseteq\mathcal V_+\cap\mathcal V_-$).
Then, the following variant of the {\em real-time KMS condition}
holds: Let $h$ be a KMS-element of $G$, with
\[h=\exp M,\ M\in\mathbf m;\]
see (\ref{eqhm}). Let
\begin{equation}
\mathcal M:=\mathrm d\pi(M)
\end{equation}
be the self-adjoint operator representing $M$ on the Hilbert
space $\mathcal H$. Then Property (\~P3), (\ref{eqptildethree}),
(\ref{eqthetavnaught}) and the theorem stated above imply that,
for arbitrary $v$ and $w$ in $\mathcal V_0$,\pagebreak
\[\langle e^{it\mathcal M}\Phi(v),\Phi(w)\rangle=
\langle J_h\Phi(w),J_he^{it\mathcal M}\Phi(v)\rangle\]
\[=\langle J_h\Phi(w),e^{it\mathcal M}J_h\Phi(v)\rangle,
\mbox{ cf.\ (\ref{eqjcommm})}\]
\[=\langle\Phi(\rho(h^{-1})\theta_rw),e^{it\mathcal M}
\Phi(\rho(h^{-1})\theta_rv)\rangle,\mbox{ cf.\ (\ref{eqsmalljh})}\]
\[=\langle e^{-\mathcal M}\Phi(\theta_rw),e^{it\mathcal M}e^{-\mathcal M}
\Phi(\theta_rv)\rangle,\mbox{ cf.\ (\ref{eqgammadef}), (\ref{eqgammal})}\]
\[=\langle e^{-i(t-i)\mathcal M}\Phi(\theta_rw),e^{-\mathcal M}
\Phi(\theta_rv)\rangle,\]
and we have used that $\mathcal M$ is self-adjoint.
The usual arguments show that
\begin{equation}\label{eqfvwone}
F_{vw}(t):=\langle e^{it\mathcal M}\Phi(v),\Phi(w)\rangle
\end{equation}
is the boundary value of a function, $F_{vw}(z)$, analytic in $z$
on the strip
\begin{equation}
\{z|0<\mathrm{Im}z<2i\},
\end{equation}
with
\[F_{vw}(t+2i)=\langle\Phi(\theta_rw),e^{it\mathcal M}\Phi(\theta_rv)\rangle\]
\begin{equation}\label{eqfvwthree}
=F_{\theta_rw\,\theta_rv}(-t).
\end{equation}

We conclude that if
\[L:=2\beta^{-1}\mathcal M\]
can be interpreted as the {\em generator of time evolution} (the
Liouvillian) in a suitably chosen frame of reference, then, apparently,
the quantum theory reconstructed in theorem~\ref{thmqft} describes
a system in {\em thermal equilibrium} at inverse temperature $\beta$.
This yields a (rather standard) imaginary-time interpretation of the
{\em Unruh-} and the {\em Hawking effects}.

Comparing equations (\ref{eqfvwone})--(\ref{eqfvwthree})
with (\ref{eqphispsi}), (\ref{eqphikms}),
we easily arrive at a formulation of the {\em connection between
spin and statistics} (SSC) in the present context.

Theorem~\ref{thmqft} and the considerations above apply to all the
examples described at the beginning of this section.\pagebreak

(i) \textbf{Minkowski space:}
\[N\equiv N_E^d=\mathbf E^d,\ N_L^d=\mathbf M^d,\]
\[G=\widetilde{SO}(d)\triangleright\!\!\!\!\times
\mathbf R^d,\ G^*=\widetilde{SO}(d-1,1)\triangleright\!\!\!\!\times
\mathbf R^d,\]
$i\mathcal M$ a boost generator (Unruh effect).

(ii) \textbf{de Sitter and $AdS$:}
(1)
\[N=S_R^d,\ N_L^d=dS_R^d,\]
\[G=\widetilde{SO}(d+1),\ G^*=\widetilde{SO}(d,1),\]
$i\mathcal M$ a boost generator {\em(``cosmic Unruh effect'')}.

(2)
\[N=\mathbf H^d,\ N_L^d=\widetilde{AdS^d},\]
\[G=\widetilde{SO}(d,1),\ G^*=\widetilde{SO}(d-1,2),\]
(``$AdS$-Unruh-effect'' [\ref{BM}]).

(iii) \textbf{Schwarzschild black hole:}
\[N=\mathbf R^2\times S^{d-2},\]
\[N_L^d=\mbox{Schwarzschild space-time},\]
\[G=\widetilde{SO}(2)\times\widetilde{SO}(d-1),\ 
G^*=\mathbf R\times\widetilde{SO}(d-1),\]
$\mathcal M$ the generator of rotations of the plane
$\mathbf R^2=U_E^2$ (i.e., $i\mathcal M\propto$ generator
of time translations, $t\mapsto t+\tau$): {\em Hawking
effect!}

We hope to present an extension of the analysis in this section
to quantum theories on more general spaces, including non-commutative
ones, elsewhere. (In this connection, note that it is really only the
symmetries $(G,K,\sigma)$ and the anti-linear involution $\theta_r$
which are important in the proof of Theorem~\ref{thmqft} and in
the remarks (1) through (3), above, but {\em not} the manifolds
$(N_E,\eta_E)$ and $(N_L,\eta_L)$!)
A particularly interesting case concerns two-dimensional conformal
field theories, where $G$ is infinite-dimensional. This case is
not covered by the results in [\ref{FOS},\ref{LM},\ref{BM}].
It would be desirable to understand it more fully.